\documentclass[a4paper,11pt]{article}
\pdfoutput=1
\usepackage{jheppub}
\usepackage{amsmath}
\usepackage{amsthm}
\usepackage{amsfonts}
\usepackage{amssymb}
\usepackage{textcomp}
\usepackage{footmisc}
\usepackage{graphicx}% Include figure files
\usepackage{bm}% bold math
\usepackage{color}
\usepackage{verbatim}
\usepackage{subcaption}
\usepackage{graphicx,color}
\usepackage{epsfig}
\usepackage{slashed}

\usepackage{ulem}

\theoremstyle{definition}

\definecolor{orange}{rgb}{1,0.5,0}
\definecolor{col1}{RGB}{153, 52, 121}
\definecolor{dgreen}{rgb}{0,0.55,0}
\definecolor{pink}{rgb}{1,0.08,0.58}

\usepackage{amsfonts,amsmath,amssymb}

\newcommand{\p}{\partial}

\newcommand{\la}{\langle}
\newcommand{\ra}{\rangle}

\newcommand{\rar}{\rightarrow}

\begin{document}

\title{
\centering \LARGE Phase relaxation and pattern formation\\ in holographic gapless charge density waves}

 \author[a]{Tomas Andrade,}

 \author[b]{Matteo Baggioli\footnote{\url{https://members.ift.uam-csic.es/matteo.baggioli}}}
 
 \author[c]{and Alexander Krikun\footnote{\url{https://orcid.org/0000-0001-8789-8703}}}

\affiliation[a]{
Departament de F{\'\i}sica Qu\`antica i Astrof\'{\i}sica, Institut de
Ci\`encies del Cosmos, Universitat de
Barcelona, \\ Mart\'{\i} i Franqu\`es 1, E-08028 Barcelona, Spain}

\affiliation[b]{Instituto de Fisica Teorica UAM/CSIC, c/Nicolas Cabrera 13-15,
Universidad Autonoma de Madrid, Cantoblanco, 28049 Madrid, Spain.
}%

\affiliation[c]{Nordita \\
KTH Royal Institute of Technology and Stockholm University \\
Roslagstullsbacken 23, SE-106 91 Stockholm, Sweden}

\emailAdd{tandrade@icc.ub.edu}
\emailAdd{matteo.baggioli@uam.es}
\emailAdd{krikun@nordita.org}

\preprint{NORDITA 2020-084, IFT-UAM/CSIC-20-122}

\abstract{
We study the dynamics of spontaneous translation symmetry breaking in holographic models in presence of weak explicit sources. 
We show that, unlike conventional gapped quantum charge density wave systems, this dynamics is well characterized by the effective time dependent Ginzburg-Landau equation, both above and below the critical temperature, which leads to a ``gapless'' algebraic pattern of metal-insulator phase transition.
In this framework we elucidate the nature of the damped Goldstone mode (the phason), which has earlier been identified in the effective hydrodynamic theory of pinned charge density wave and observed in holographic homogeneous lattice models. 
We follow the motion of the quasinormal modes across the dynamical phase transition
in models with either periodic inhomogeneous or helical homogeneous spatial structures,
showing that the phase relaxation rate is continuous at the critical temperature. 
Moreover, we find that the qualitative low-energy dynamics of the broken phase is universal, insensitive to the precise pattern of translation symmetry breaking,
and therefore applies to homogeneous models as well.}

\maketitle

\section{Introduction}
Interacting quantum matter exhibits a large variety of long-range orders in which spatial symmetries are broken. These patterns are ubiquitous within the phase diagram of strongly correlated materials such as high temperature superconducting cuprates \cite{RevModPhys.75.1201,doi:10.1080/00018730903122242,berg2009striped,Chen_2016}. Charge density waves (CDW) represent the simplest example among them, where translational invariance is spontaneously broken by the dynamical generation of a modulated periodic charge density distribution \cite{RevModPhys.60.1129}. The interest around this topic has been further motivated by the observation that the interplay of explicit and spontaneous breaking of translations could be the key behind the DC and optical electric transport properties of bad metals and strange metals \cite{Delacretaz:2016ivq,Amoretti:2017axe} as well as the strongly correlated Mott insulating states \cite{Mott}. Moreover, the dynamics and the collective modes through a symmetry breaking transition are under intense experimental investigation \cite{Yusupov2010,PhysRevB.89.045106,Thomson2017}.

The holographic approach to strongly correlated quantum matter \cite{Zaanen:2015oix,Hartnoll:2016apf,Baggioli:2019rrs} has proven to be quite useful in describing the physics of translational symmetry breaking (TSB), both explicit and spontaneous. Its classical gravitational dynamics allows one to build a variety of models with different complexity, which provide a useful theoretical framework for studying various aspects of TSB. The spectrum of available models includes those with spontaneous symmetry breaking through periodic structures \cite{Donos:2011bh,Donos:2013wia,Donos:2013gda,Withers:2013loa,Withers:2013kva}, or in a helical pattern \cite{Nakamura:2009tf,Ooguri:2010kt,Donos:2012wi}, as well as explicit TSB by periodic external potentials \cite{Horowitz:2012ky,Horowitz:2012gs,Rangamani:2015hka,Donos:2014yya}, helical sources \cite{Donos:2012js,Donos:2014oha}, or completely homogeneous \cite{Andrade:2013gsa,Donos:2013eha,Baggioli:2014roa,Alberte:2015isw}. Noteworthy, these latter models, the holographic homogeneous lattices, treat TSB in a way similar to effective field theory for solids \cite{Nicolis:2015sra} and don't involve the actual periodic patterns. 

More recently, the physics of the weakly pinned spontaneous spatial structures has been addressed within the holographic framework. It has been shown that in presence of explicit sources the spontaneously broken phases display a softly gapped pseudo-Goldstone mode \cite{Andrade:2017cnc,Alberte:2017cch, Amoretti:2019kuf}. 
Related to this, the formation of weakly pinned spontaneous structure leads to a metal-insulator phase transition \cite{Mott,Andrade:2018gqk}, as one would expect for conventional pinned charge density wave and Mott insulating states \cite{RevModPhys.60.1129}. 
It turns out however that the detailed features of this transition don't quite follow the conventional pattern for gapped quantum systems \cite{RevModPhys.60.1129}.
Instead the transport in holographic models with weakly pinned CDW is well described by the effective hydrodynamic approach developed in \cite{Delacretaz:2016ivq,Delacretaz:2017zxd}.
This has been demonstrated in numerous examples in holographic homogeneous setups \cite{Amoretti:2018tzw,Ammon:2019wci,Donos:2019hpp,Amoretti:2019cef,Andrade:2018gqk}%
\footnote{Note an interesting disagreement found in \cite{Ammon:2019apj} and further resolved in \cite{Armas:2019sbe,Armas:2020bmo,Ammon:2020xyv,Baggioli:2020edn}.}
 and, as we show below, also holds well in the case of inhomogeneous holographic lattices. 

In the models under consideration the DC conductivity remains finite in the broken phase, unlike the standard result of  \cite{RevModPhys.60.1129}, where the Peierls instability leads to an opening of the gap in the spectrum of charge carriers which directly implies insulator-like exponential suppression of conductivity.
Holography even allows for the presence of metallic broken states \cite{Amoretti:2017axe} which were suggested to be relevant for the phenomenology of bad metals \cite{Delacretaz:2016ivq}. 
From the technical point of view, this distinct behavior is a direct consequence of the presence of additional transport coefficients which are allowed in holographic models due to the absence of Galilean invariance (which would otherwise be a direct consequence of the  quasiparticle approach of \cite{RevModPhys.60.1129}): the incoherent conductivity term $\sigma_{\mathrm{inc}}$, the dissipative parameter $\gamma$ and the new phase relaxation term $\Omega$. 
More explicitly, the hydrodynamic treatment \cite{Delacretaz:2016ivq,Delacretaz:2017zxd} gives the following expression for the AC conductivity
\begin{equation}
\label{AC model}
\sigma(\omega) = \sigma_{\mathrm{inc}} + \frac{\frac{\rho^2}{\chi_{\pi\pi}}\left(\Omega- i \omega\right) - \omega_0^2\,\gamma\,[2\,\rho+\gamma \chi_{\pi\pi}(\Gamma-i \omega)]}{(\Gamma - i \omega)(\Omega - i \omega) + \omega_0^2}\,.
\end{equation}
Here, $\chi_{\pi\pi}\,=\,\epsilon+\langle T^{xx}\rangle$\footnote{Here and everywhere below we consider translation symmetry breaking along the $x$-axis and measure the conductivity in this direction.} is the static momentum susceptibility, $\Gamma$ -- the momentum relaxation rate, $\omega_0$ is the pinning frequency which controls the mass of the pseudo-Goldstone. 
$\sigma_{\mathrm{inc}}$~is a part of conductivity which corresponds to a movement of charges with a net zero momentum, 
%similar to creation of particle-antiparticle pairs, 
which can occur in systems where Galilean invariance is broken (e.g. systems with no quasiparticles) \cite{Davison:2015bea,Davison:2015taa}. The latter is responsible for finite residual conductivity in the holographic systems where coherent transport is suppressed due to explicit mechanisms of~TSB: impurities, crystal lattice or disorder. In this case $\Gamma$ is large, $\omega_0 = 0$ due to the absence of spontaneous order, and therefore $\sigma_{DC} \approx \sigma_{\mathrm{inc}}$. This contribution means the absence of disorder-driven insulating states with gapped conductivity in holographic models~\cite{Grozdanov:2015qia}\footnote{Notice the counterexamples in \cite{Baggioli:2016oqk,An:2020tkn}.}, where the conductivity drops as a power law instead $\rho \sim \sigma_{\mathrm{inc}}^{-1} \sim T^{-\alpha}$ \cite{Donos:2012js,Donos:2014uba, Andrade:2018gqk, Mott}.

In presence of the spontaneous structure, however, $\omega_0$ is nonzero in \eqref{AC model} and the DC conductivity gets an extra contribution due to finite $\Omega$, which is the focus of our present work and which defines the pattern of how exactly the conductivity drops across the metal-insulator phase transition.
The ``phase relaxation rate'' $\Omega$ characterizes the lifetime of the Goldstone mode (phason) in the system. Originally, in \cite{Delacretaz:2017zxd}, this decoherence of Goldstone has been attributed to the effect of topological defects (e.g. dislocations or disclinations) in the spontaneous spatial structure, which would destroy the order at large distances. However, it has been shown that the imprints of $\Omega$ appear also in the homogeneous models, where no spatial structure of the order parameter is present, and therefore no topological defects could be introduced. Moreover, in several examples \cite{Amoretti:2018tzw,Ammon:2019wci,Donos:2019hpp,Amoretti:2019cef,Andrade:2018gqk}, it has been observed that the phase relaxation rate is proportional to the scale of explicit symmetry breaking and therefore can not be a feature of the spontaneous structure only. Notably, up to now the phase relaxation rate has only been studied in homogeneous models without dynamical phase transition. So it hasn't been completely clear whether finite $\Omega$ is an artifact of these simplified models, or a misconception in the hydrodynamic approach.

\textbf{In this work,} we show that the damped Goldstone mode, characterized by finite phase relaxation rate $\Omega$ is a generic feature of dynamical spontaneous symmetry breaking in presence of weak explicit sources, and it can be described via a dissipative time dependent Ginzburg-Landau equation (TDGL). More precisely, we show that it appears due to a splitting of the unstable modes, which drive the phase transition at critical temperature. This splitting is caused by an explicit symmetry breaking term which introduces a finite relaxation time for the fluctuations of the phase of the spontaneous structure.
This effect can be already seen in the classical model as simple as $U(1)$ symmetry breaking with $\phi^4$ potential, as well as in the Ginzburg-Landau treatment of pattern formation in classical systems \cite{PhysRevLett.56.724}. 

We demonstrate that the holographic models, which break translations dynamically, follow the same pattern, both in case of inhomogeneous and homogeneous (helical) TSB patterns. 
In order to get optimal control over $\Omega$ in these models we study the quasinormal mode spectrum and the AC conductivity in the immediate vicinity of the dynamical phase transition. In this regime, when $\omega_0^2 \ll \Gamma \Omega$, the expression \eqref{AC model} has two purely imaginary poles in the lower half-plane, which at $T_c$ correspond exactly to $\Gamma$ and $\Omega$. We show that this feature is indeed present in the models under consideration, where $\Omega$ is finite and the it corresponds exactly to the damped Goldstone mode predicted by the TDGL effective treatment. 

\textbf{Our main finding} is therefore, that the damped Goldstone mode, $\Omega$, is not an artifact of erroneous or oversimplified treatments, but rather a very generic feature of the holographic models, which is deeply rooted in the fact that holographic systems are characterized by the dissipative TDGL dynamics even in the ordered phase. The difference with the conventional quantum systems is then in the fact that the latter don't admit the TDGL treatment due to the gap in the quasiparticle spectrum, while in the former quasiparticles are absent and dissipative processes are never suppressed. 
This effect reflects the general feature of the holographic models of strange metals -- the spectral density is always finite due to the presence of \textit{quantum critical continuum} and therefore none of the signatures of the true gapped quantum systems appear \cite{Krikun:2018agd}.

Moreover, we observe that the dynamics of phase relaxation is continuous across the critical temperature and universal in the broken phase. This universality relies on the fact that local structure of the spatial pattern is irrelevant at large distance \cite{Nicolis:2015sra} and that the dynamics in the broken phase is surprisingly insensitive to the presence of a proper dynamical instability and to its nature. In result, even a minimal TSB model like the axions-like one \cite{Baggioli:2014roa} is able to reproduce all the fundamental features of the broken phase in presence of pinning. 

Finally, we explain in detail the role of the phase relaxation $\Omega$ and the pinning frequency $\omega_0$ in the algebraic metal-insulator transition happening at the critical temperature. 
Given the contrast with the standard exponentially gapped CDWs, and in analogy with gapless superconductors, which are also well described by TDGL \cite{gor1996generalization}, we label our setup as \textbf{holographic ``gapless'' charge density waves}.
%\\[0.2cm]

The paper is organized as follows: We start in Sec.\,\ref{sec:model_TSB} by introducing the TDGL framework with explicit symmetry breaking sources as applied to $U(1)$ $\phi^4$ theory and theory of classical pattern formation. In Sec.\,\ref{sec:periodic} we discuss the holographic model with dynamic formation of periodic CDW and show that its spectrum of quasinormal modes follows the prediction of TDGL quite precisely. We study the effect of the damped Goldstone in the conductivity in Sec.\,\ref{sec:MIT} and show that it agrees perfectly with hydrodynamic treatment \eqref{AC model}. In Sec.\,\ref{sec:helical} we demonstrate that the homogeneous helical model behaves similarly despite the fact that is lacks true periodicity. Furthermore, in Sec.\,\ref{sec:homogeneous} we discuss why the earlier results obtained in homogeneous models without dynamical phase transition fit well into the same pattern. We conclude in Sec.\,\ref{sec:conclusion}. The three Appendices are devoted to the technical details and extra datasets for the periodic and helical models, as well as the wider treatment of TDGL theory in presence of explicit sources with various commensurability fractions.

\section{\label{sec:model_TSB}Time-dependent Ginzburg-Landau equation and pattern formation}
%\subsection{Phase transitions in presence of explicit source: U(1) case}

Let us first consider the simplest model for spontaneous breaking of global U(1) symmetry described by a complex scalar field $\Phi$ with Ginzburg-Landau free energy
\begin{gather}
\label{equ:GL}
\mathcal{F} = \alpha |\Phi|^2 + \frac{\beta}{2} |\Phi|^4, \qquad \alpha \sim (T-T^0_c), \qquad \beta>0 \,,\\
\notag
\Phi \equiv \varphi e^{i \vartheta}\,.
\end{gather}
Depending on the sign of parameter $\alpha$ it has a minimum either at $|\Phi| = 0$ for $\alpha>0$ -- normal phase, or at $|\Phi| = \varphi_0 = \sqrt{-\alpha/\beta}$ for $\alpha<0$ -- broken phase. The phase transition happens at critical temperature $T^0_c$.
Notice that the global U(1) symmetry leads to a degenerate set of ground states in the broken phase $\Phi_0 = \varphi_0 \,e^{i \vartheta_0}$, which are  parametrized by the complex phase $\vartheta_0 \in [0,2\pi)$.
% \begin{equation}
% \mbox{Broken phase states:} \qquad \Phi_0 = \varphi_0 \,e^{i \vartheta_0}, \qquad \vartheta_0 \in [0,2\pi).
% \end{equation}
The transitions between them cost no energy and therefore are mediated perturbatively by a massless Goldstone mode, which is nothing but a phase shift: $\vartheta_0 \rar \vartheta_0 + \delta \vartheta$.

\begin{figure}
\center
\raisebox{-0.5\height}{\includegraphics[width=0.6\linewidth]{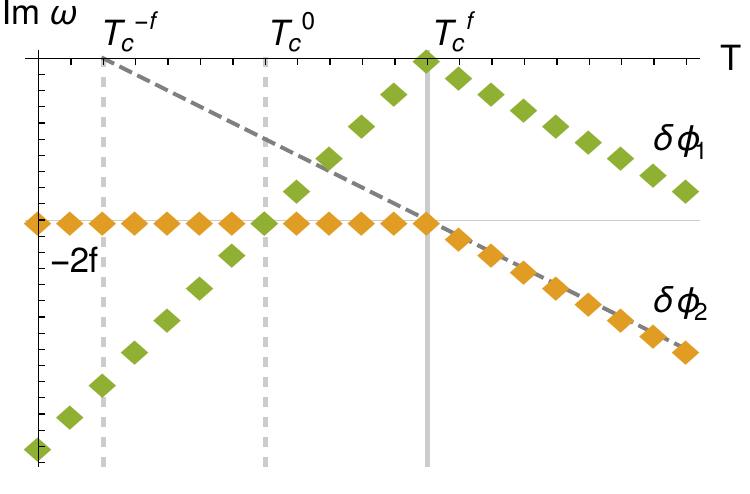}} 
\qquad
\raisebox{-0.5\height}{\includegraphics[width=0.25\linewidth]{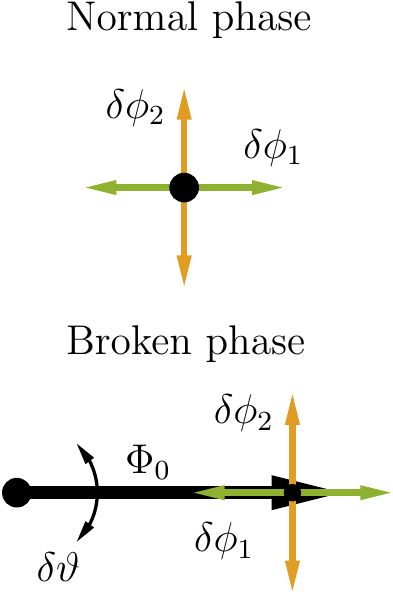}}
\caption{\label{fig:U1QNMs} \label{fig:U1cartoon} 
\small
\textbf{Right panel:} The roles of the fluctuations of real $\delta \phi_1$  and imaginary  $\delta \phi_2$ modes of the complex scalar \eqref{equ:ImRe_phi}. In the normal state they are degenerate and related by $U(1)$ rotation. In the broken state characterized by real order parameter $\Phi_0$ the mode $\delta \phi_1$ corresponds to the fluctuation of the order parameter, while  mode $\delta \phi_2$ is the fluctuation of its phase -- the Goldstone mode. \\
\textbf{Left panel:} 
 The dynamics of the $\delta \phi_{1,2}$ modes across the $U(1)$ phase transition coming from the time dependent Ginzburg-Landau analysis \eqref{equ:def_time_GL} in presence of explicit symmetry breaking term $f$ . The critical temperature is shifted by the external force ($T_c^0 \rar T_c^f$). Above $T_c^f$ the degeneracy between the two instability driving modes is resolved \eqref{equ:deformed_modes_above_Tc}. Below $T_c^f$ the phase mode turns into damped Goldstone \eqref{equ:def_modes_below_Tc}.
 }
\end{figure}

It is instructive to consider a parametrization of the complex scalar $\Phi$ in terms of its real and imaginary components
\begin{equation}
\label{equ:ImRe_phi}
\Phi \equiv \phi_1 + i \phi_2\,.
\end{equation}
We can always pick the direction of the ground state in the broken phase along the real axis ($\phi_1 = \varphi_0, \, \phi_2 = 0$). Then the Goldstone mode is completely contained in the $\delta \phi_2$ fluctuation, i.e. $\vartheta_0 \rar \vartheta_0 + \delta \vartheta$ is equivalent to $\phi_2 \rar \phi_2 + \varphi_0\, \delta \vartheta$, see cartoon on Fig.\,\ref{fig:U1cartoon}.
% \begin{align}
% \label{equ:broken_phase}
% \mbox{Ground state:} \qquad \Phi_0 &= \varphi_0 \\
% \phi_1 &= \varphi_0, \qquad \phi_2 = 0\,;\\
% \mbox{Goldstone mode:} \qquad \Phi_0 &\rar \Phi_0\, e^{i \delta \vartheta} \\
%  \phi_2 &\rar \phi_2 + \varphi_0\, \delta \vartheta\,.
% \end{align}

In order to analyze the stability of a particular phase one can make use of the time-dependent Ginzburg-Landau equation
\begin{equation}
\label{equ:time_GL}
\p_t \Phi = -\frac{\delta \mathcal{F}}{\delta \Phi*} = - \alpha \Phi - \beta |\Phi|^2 \Phi.
\end{equation}
Linearizing this equation around the normal phase $\phi_1 = \phi_2 = 0$, we obtain the evolution equations for fluctuations $\delta \phi_{1,2}$, which give two degenerate exponentially decaying modes above $T^0_c$ ($\alpha>0$):
\begin{align}
\label{equ:instabilities}
\p_t \delta \phi_{1,2} &= - \alpha \delta \phi_{1,2} \,, & \delta \phi_{1,2} &\sim e^{-\alpha t}. 
\end{align}
The fact that the ground state is stable is, of course, expected. The TDGL equation \eqref{equ:time_GL} for linear perturbations equates their time derivatives to the minus second derivative of the free energy \eqref{equ:GL}, evaluated in the ground state. Since the latter is the true minimum, the second derivative is positive.
It is worth noticing that as one approaches $T^0_c$, $\alpha$ decreases and eventually changes sign. Then the modes \eqref{equ:instabilities} become unstable (i.e. exponentially growing) and mediate the phase transition towards a new, broken phase.

The stability of the broken phase can be analyzed similarly. Expanding around the new ground state $\phi_1 = \varphi_0 = \sqrt{- \alpha/\beta}, \phi_2 = 0$, we get ($\alpha<0$)
\begin{align}
\p_t \delta \phi_1 &= - (\alpha + 3 \beta \varphi_0^2) \delta \phi_1(t) = 2 \alpha \delta \phi_1  & \delta \phi_{1} &\sim e^{2 \alpha t}, \\
\notag
\p_t \delta \phi_2 &= - (\alpha + \beta \varphi_0^2) \delta \phi_2(t) = 0 &  \delta \phi_{2} &\sim const
\end{align}
Noteworthy, the $\delta \phi_2$ mode does not decay in the broken phase. It is a Goldstone mode and is protected by the symmetry. On the other hand the fluctuations of $\delta \phi_1$ decay and the broken phase is stable.

Let us move on and consider how the picture is modified when one introduces an explicit symmetry breaking term in the free energy \eqref{equ:GL}. Consider the deformation
\begin{equation}
\label{equ:GL_deformed}
\mathcal{F}_f = \alpha |\Phi|^2 + \frac{\beta}{2} |\Phi|^4 - \frac{f^*}{2} \Phi^2 - \frac{f}{2} {\Phi^*}^2.
\end{equation}
The complex phase of $f$ sets the preferred direction of the order parameter. Without loss of generality we can choose $f$ to be real and positive.
Now the free energy is no more invariant under the U(1) symmetry transformation. The continuous shifts in phase $\vartheta$ are now broken down to a discrete subgroup $\vartheta \rar \vartheta + \pi$, since $\mathcal{F}$ is still insensitive to the change of sign of $\Phi$. 

At large enough positive $\alpha$, the ground state of the deformed system is still simply in the normal phase with $|\Phi| = 0$. Let's analyze its stability right away. The time dependent GL equation reads
\begin{equation}
\label{equ:def_time_GL}
\p_t \Phi = -\frac{\delta \mathcal{F}}{\delta \Phi*} = - \alpha \Phi - \beta |\Phi|^2 \Phi + f \Phi^*.
\end{equation}
In the normal phase this gives for fluctuations
\begin{align}
\label{equ:deformed_modes_above_Tc}
\p_t \delta \phi_1 &= - (\alpha - f) \delta \phi_1, & \delta \phi_{1}(t) &\sim e^{-(\alpha - f) t}, \\
\notag
\p_t \delta \phi_2 &= - (\alpha + f) \delta \phi_2, & \delta \phi_{2}(t) &\sim e^{- (\alpha+f) t}. 
\end{align}
We see immediately that the explicit breaking lifts a degeneracy between real and imaginary components of the complex scalar; therefore above $T^0_c$ there will be two exponentially decaying modes in the spectrum, see Fig.\,\ref{fig:U1QNMs}. 
Note also that the $\delta \phi_1$ mode now becomes unstable at a temperature $T^f_c$ higher then $T^0_c$, when $[\alpha (T^f_c) - f]$ becomes negative. Therefore, the explicit breaking changes the temperature of the phase transition. Another important observation here is that at the moment when $\delta \phi_1$ becomes unstable, the other mode $\delta \phi_2$ is still decaying with the rate $\delta \phi_2 \sim e^{-2 f t}$.

The instability at $T^f_c$ leads to a ground state with finite $\phi_1$, defined by the minimum of the free energy \eqref{equ:GL_deformed}: $\Phi_0^f = \varphi^f = \sqrt{- (\alpha - f)/\beta}$, (note that no freedom in choosing a complex phase is present).
It is worth mentioning that despite the fact that $U(1)$ symmetry has been explicitly broken in the system, the phase transition at $T^f_c$ is still of second order since it breaks the remaining discrete symmetry $\mathbb{Z}_2: \Phi \rar - \Phi$ spontaneously.  
Note also that when the temperature is further lowered below $T_c^{-f}: \alpha(T_c^{-f}) + f = 0$ the free energy has also another extremum at purely imaginary $\Phi_0^{-f} = i \sqrt{- (\alpha + f)/\beta}$, but this one is not a true minimum and is unstable.

Consider now the stability of the broken phase formed at $T^f_c$. We linearize the deformed time dependent GL equation \eqref{equ:def_time_GL} about the solution $\phi_1 = \varphi^f, \phi_2 = 0$ and observe the two modes:
\begin{align}
\label{equ:def_modes_below_Tc}
\p_t \delta \phi_1 &= - \left( (\alpha - f) + 3 \beta (\varphi^f)^2 \right) \delta \phi_1 = 2 (\alpha - f) \delta \phi_1,  & \delta \phi_{1}(t) &\sim e^{2 (\alpha-f) t}, \\
\notag
\p_t \delta \phi_2 &= - \left( (\alpha + f) + \beta (\varphi^f)^2 \right) \delta \phi_2 = - 2 f \delta \phi_2, &  \delta \phi_{2}(t) &\sim  e^{- 2 f t}.
\end{align}
This is the key result of our treatment: in the broken phase the explicit symmetry breaking leads to a damping of ``would be'' Goldstone mode with the rate $\Omega = -2f$. Therefore the damped Goldstone appears in the spectrum below the phase transition, see Fig.\,\ref{fig:U1QNMs}.

\begin{figure}
\raisebox{-0.5\height}{\includegraphics[width=0.5 \linewidth]{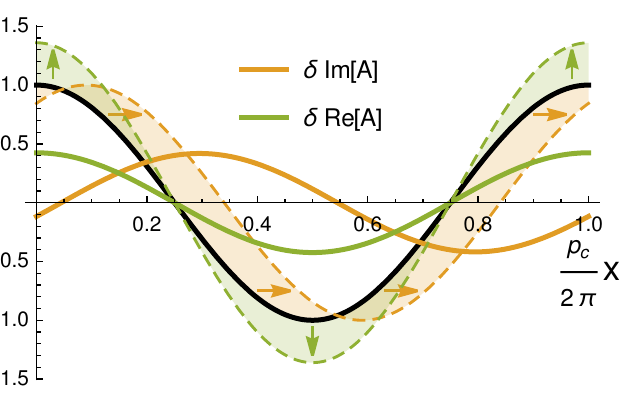}}
\raisebox{-0.5\height}{\includegraphics[width=0.5 \linewidth]{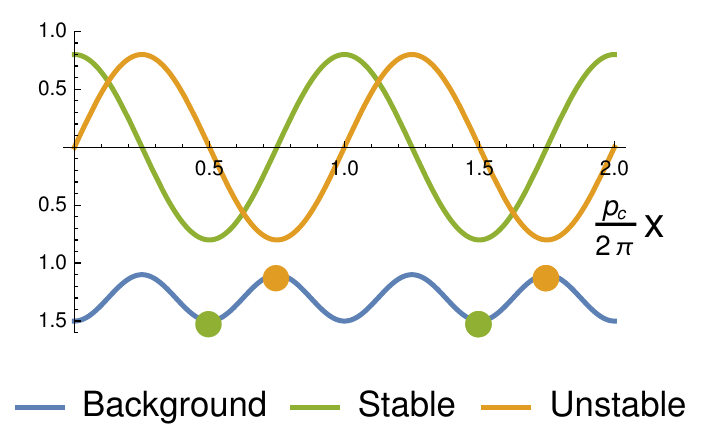}}
\caption{\label{fig:toy_model_periodic}
\small
\textbf{Left panel:} Amplitude and phase modes in case of translation breaking due to periodic spontaneous structure. Black line is the order parameter. Shifts along the $x$-axis correspond to imaginary part of amplitude $A$ \eqref{equ:ImReA}, yellow line, variation of amplitude corresponds to real part, green line. \\
\textbf{Right panel:} Stable and unstable ground states, in presence of explicit symmetry breaking potential (blue line). The two states are shifted by a quarter period. Minima of the stable state (purely real $A$) correspond to minima of the explicit potential, while the minima of the unstable state (purely imaginary $A$), to the maxima.}
\end{figure}

The discussion above can almost literally be translated to the case, where instead of U(1) global symmetry one considers the \textbf{spontaneous breaking of translations} by an emergent spatial superstructure. The relevant treatment can be found in a vast literature addressing the theory of \textit{pattern formation}, see \cite{PhysRevLett.56.724,cross1993pattern}. For consistency, we summarize the treatment of \cite{PhysRevLett.56.724} here. The formation of a periodic structure is driven by an unstable mode with finite momentum $p_c$ -- the critical wave-vector. One can build a Ginzburg-Landau like effective theory of such a transition by considering the ansatz for a given physical observable. The simplest case of static pattern without defects is described by
\begin{equation}
\label{eq:A_ansatz}
    \Psi(t,x)\,=\,\Psi_0\,+\,A(t)\, e^{i\,p_c\,x}\,+\,A(t)^*\, e^{-i\,p_c\,x}\,+\,\dots
\end{equation}
with a complex valued amplitude $A(t)$ and $\Psi_0$ denoting the background homogeneous value in the normal phase. In order to describe the instability, one can write down an equation for the amplitude $A$, known as real Ginzburg-Landau equation or simply \textit{amplitude equation}:
\begin{equation}
\label{equ:amplitude equation}
    \partial_t A\,= - \alpha A - \beta |A|^2\,A,
\end{equation}
with, again, the sign of parameter $\alpha$ governing the phase transition.
This is, of course, identical to the time dependent Ginzburg-Landau equation of the U(1) case \eqref{equ:time_GL}. Let us only elaborate a bit more on the physical meaning of $A$. On one hand, introducing its modulus and phase we get from \eqref{eq:A_ansatz}
\begin{equation}
\Psi(t,x)\,=\,\Psi_0\,+ |A(t)| \cos(p_c x + \vartheta), \qquad A(t) \equiv |A(t)|\, e^{i \vartheta(t)}.
\end{equation}
In other words, the phase of $A$ accounts for the position of the spontaneous structure. Similarly to the $U(1)$ example discussed above, in case the system is translationally invariant, the phase of the broken state is arbitrary and therefore the change of $\vartheta$ is again a Goldstone mode. It is related to the shift of the full spontaneous structure along the $x$-coordinate and is usually labeled as \textit{the phason}. In particular
\begin{equation}
\label{equ:shifts}
\vartheta \rar \vartheta + \Sigma 
\qquad
\Leftrightarrow 
\qquad 
x \rar x + \Sigma/p_c.
\end{equation}
If we rewrite \eqref{eq:A_ansatz} as, 
\begin{equation}
\label{equ:ImReA}
\Psi(t,x)\,=\,\Psi_0\,+\, \mathrm{Re}[A(t)]  \cos(p_c x) + \mathrm{Im}[A(t)] \sin(p_c x),
\end{equation}
we see that the real and imaginary parts of $A(t)$ correspond to the amplitude of the patterns, shifted by a quarter of the period. Notably, again, the fluctuations of the real and imaginary parts of $A$, around the ground state described by i.e. $\cos(p_c x)$-pattern correspond to the fluctuations of the amplitude or the phase of the order parameter, see Fig.\,\ref{fig:toy_model_periodic}:
\begin{align}
%\qquad A_0  \cos(p_c x) &\rar 
&A_0 \cos(p_c x + \delta \vartheta),  & &\Leftrightarrow & \delta \mathrm{Im}[A] &= A_0 \delta \vartheta \\
%\qquad A_0  \cos(p_c x) &\rar 
\notag
&(A_0 + \delta A) \cos(p_c x), & &\Leftrightarrow &  \delta \mathrm{Re}[A] &= \delta A
% &= A(t) \cos(q_c x) + \delta \vartheta A(t) \sin(q_c x)\\
%  \delta \mathrm{Im} A  &=   A(t) \delta \vartheta, \qquad  \delta \mathrm{Re} A  =  0.
\end{align}
In other words, the imaginary part of A, which multiplies a $\sin(p_c x)$ pattern in \eqref{equ:ImReA} represents the phason mode around the $\cos(p_c x)$ ground state, see Fig.\,\ref{fig:toy_model_periodic}. 

Let's now consider what happens if we introduce a periodic explicit potential with wave-vector $k$, commensurate with $p_c$ \cite{PhysRevLett.56.724}. In this case the continuous translation symmetry, reflected as the phase rotations in $A$ \eqref{equ:shifts} is broken down to a discrete set
\begin{equation}
x \rar x + \frac{2\pi}{k} n, \qquad \vartheta \rar \vartheta + 2 \pi \frac{p_c}{k} n, \qquad n \in \mathbb{Z}.
\end{equation}
In here we'll focus only on the case $k = 2 p_c$, then the remaining discrete symmetry is $\mathbb{Z}_2$ and corresponds to the change of phase by $\pi$ (or simply the change of sign of field $A$). The extra term representing the explicit potential, which respects this $\mathbb{Z}_2$ symmetry, can be added to the amplitude equation \eqref{equ:amplitude equation}, which then reads
\begin{equation}
\label{equ:amplitude equation_deformed}
    \partial_t A\,= - \alpha A - \beta |A|^2\,A + f A^*.
\end{equation}
Here we recover literally the analogy with the deformed time dependent Ginzburg-Landau equation \eqref{equ:def_time_GL}. All the same consequences follow, including the existence of two separate purely damped modes above critical temperature and the damped Goldstone mode in the broken phase, see Fig.\,\ref{fig:U1cartoon}. Let us only recap again the physical meaning of these modes.

The two unstable modes, parametrized by real and imaginary parts of $A$, correspond to the two possible patterns of spontaneous structure, aligned with the ``maxima'' or with the ``minima'' of the background potential, respectively, see Fig.~\ref{fig:toy_model_periodic}. Depending on the features of the model, one of them will drive the phase transition at $T^f_c$ and lead to formation of the ground state. In the broken phase, the fluctuations of the other $\pi/4$-shifted mode will play the role of the damped Goldstone.

Before we move to the concrete examples of this dynamics, provided by holographic models, let's introduce one more ingredient to the system -- the electric charge and associated current. If the system has a finite, homogeneously distributed charge density in a normal state (which can, for instance, be accounted by the leading term in \eqref{eq:A_ansatz}), then in case translations are not broken explicitly
the homogeneous movement of this charge density will mediate infinite conductivity due to conservation of its momentum. However, if momentum is not conserved, then the homogeneous movement of charge density is relaxed and one obtains a Drude peak in the conductivity, related to a dissipative mode at $\omega=-i \Gamma$ in the spectrum of the fluctuations. Here $\Gamma$ is the momentum relaxation rate, c.f. \eqref{AC model}.

When the spontaneous structure is formed, it possesses a finite charge. Therefore its shifts, accounted for by the damped Goldstone mode, contribute to the electric current. 
Therefore in the broken phase the current two point function (conductivity) acquires an extra pole $\omega=-i \Omega$ in the spectrum, related to the damped shifts of the spontaneous structure. This discussion goes completely in lines with the effective hydrodynamic treatment of \cite{Delacretaz:2017zxd}.

\section{\label{sec:periodic}The holographic periodic charge density wave}

Consider now a concrete holographic model, which displays the dynamical spontaneous translational symmetry breaking. We will start from the model which develops a periodic pattern of staggered transverse currents coupled to a modulation of the pseudoscalar field, which was introduced in \cite{Donos:2011bh, Donos:2013wia,Withers:2013kva, Withers:2013loa}, and add the explicit periodic ionic lattice by modulating the chemical potential, as in \cite{Horowitz:2012gs,Donos:2014yya,Rangamani:2015hka}. This system has been studied extensively in \cite{Mott,Krikun:2017cyw,Andrade:2017leb}. The model includes Einstein-Maxwell theory with an axion field coupled to a gauge $\theta$-term in 4-dimensional bulk spacetime (with holographic boundary at $z=0$):
\begin{equation}\label{S_0}
  S = \int d^4 x \sqrt{- g} \left( R - \frac{1}{2} (\partial \psi)^2 - \frac{\tau(\psi)}{4} F^2 - V(\psi) \right)
   - \frac{1}{2} \int {\theta}(\psi) F \wedge F
\end{equation}
Here $F = d \mathcal{A}$ is the field strength of the $U(1)$ gauge field $\mathcal{A}_\mu$. Following 
\cite{Mott,Donos:2011bh, Withers:2013kva, Withers:2013loa, Donos:2013wia}, the couplings are chosen to be 
\begin{gather}
\label{equ:potentials}
  V(\psi) \equiv 2 \Lambda +W(\psi) = - 6 \cosh (\psi /\sqrt{3}) , \quad \\
  \notag
   \tau(\psi) = {\rm sech} (\sqrt{3} \psi), \quad \theta(\psi) = \frac{c_1}{6 \sqrt{2}} \tanh(\sqrt{3} \psi),
\end{gather}
Note that in these conventions the cosmological constant is $\Lambda = - 3$ and the mass of the scalar is $m^2 = -2$. In what follows we pick the value $c_1=17$ used earlier in \cite{Mott,Krikun:2017cyw}, so we are in the identical setup and refer to those works for more technical details.
The explicit translation symmetry breaking is introduced by a modulation of the chemical potential which describes the ionic lattice:
\begin{equation}\label{eq:mu x}
 \mathcal{A}_t(x) \big|_{z\rar 0} = \mu(x) \equiv \mu_0 (1 + A \cos(k x) ).
\end{equation}

As it was shown in \cite{Donos:2011bh, Withers:2013kva, Withers:2013loa, Donos:2013wia}, due to the $\theta$-coupling this model in absence of the explicit potential develops an instability at low temperature evolving into the spatially modulated ground state, which breaks translations spontaneously and features oscillating diamagnetic currents $J_y \sim \cos(x p_c + \vartheta)$ and axion field $\psi \sim \sin(x p_c + \vartheta)$. In this broken state the modulated charge density (CDW) arises as well with the wave vector $2 p_c$. When introducing the external TSB, one can chose the modulation of the chemical potential exactly commensurate with this CDW, which leads to a situation described above: the explicit periodic force has twice a wave-vector of the spontaneous structure \cite{Mott}.

The phase transition in the model is driven by a marginal mode ($\omega = 0$) appearing in the spectrum at critical temperature. As it has been studied in \cite{Donos:2011bh,Andrade:2017leb}, this marginal mode includes the fields
\begin{equation}\label{eq:marginal modes}
  \delta \psi = \delta \psi^p(z, x) e^{i p_c x}, \qquad \delta A_y = \delta {A_y}^p(z, x) i e^{i p_c x}, 
  \qquad \delta Q_{ty} = \delta {Q_{ty}}^p(z, x) e^{i p_c x}.
\end{equation}
In \cite{Andrade:2017leb} is has been also noted that the external potential affects the critical temperature of the phase transition, in qualitative agreement with our GL model treatment \eqref{equ:deformed_modes_above_Tc}. Moreover, the existence of the subleading instability (the one at $T_c^{-f}$, see Fig.\,\ref{fig:U1cartoon}) has also been observed already in \cite{Andrade:2017leb} (see e.f. top right panel of Fig.\,4 there)\footnote{More precisely, the only difference from the case we consider now is that in \cite{Andrade:2017leb} the wave-vectors of the explicit and spontaneous structures were not initially tuned to be commensurate. Therefore the commensurability and the splitting of two critical temperatures was happening only at finite amplitude of the explicit potential, when the spontaneous structure is forced to acquire a commensurate wave-vector. In the situation considered here, the periodicity of the explicit lattice is commensurate to the spontaneous wave-vector, therefore the splitting happens at arbitrarily small amplitude. In terms of the toy-model, discussed in \cite{Andrade:2017leb}, the two bell curves corresponding to the two possible unstable modes intersect due to the interaction with the commensurate lattice in such a way that the corresponding ``umklapp''-gap opens precisely at the tip of the coinciding bells, therefore leading to the split of the critical temperature, which we observe here.}.
Let us now expand the study of \cite{Andrade:2017leb} and explore the spectrum of all quasinormal modes (QNMs), with $\omega \neq 0$, in the system as its temperature goes across the phase transition.

\begin{figure}
\centering
\raisebox{-0.5\height}{\includegraphics[width=0.63 \linewidth]{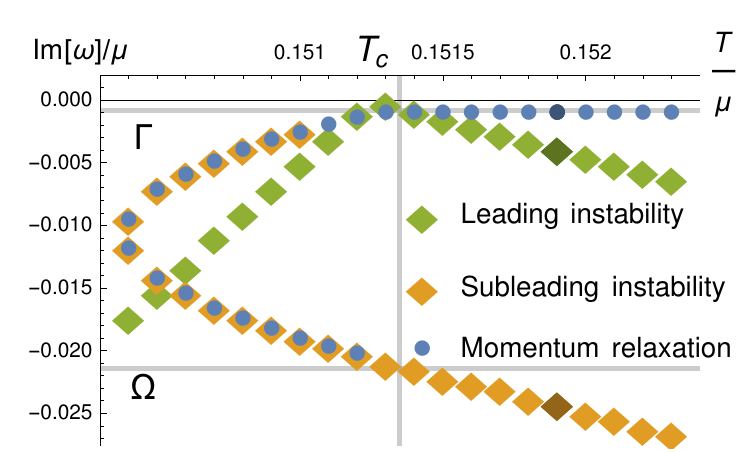}}
\hfill
\raisebox{-0.5\height}{\includegraphics[width=0.33 \linewidth]{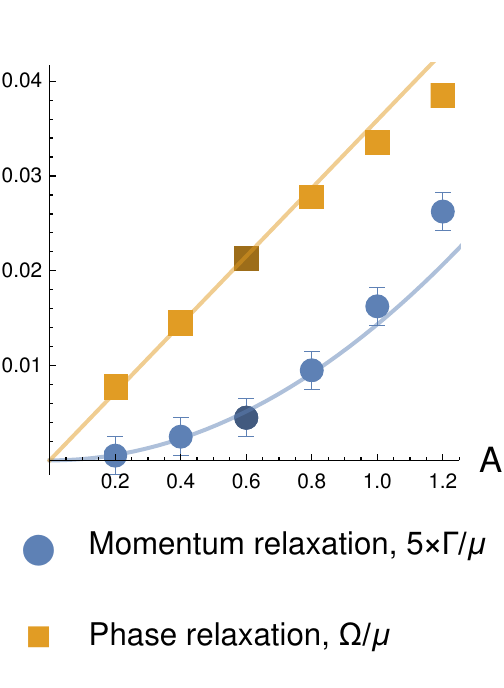}}
\caption{\label{fig:periodic_modes}\label{fig:omega_A}
\small
\textbf{Left panel:}
Spectrum of the purely imaginary quasinormal modes in holographic model with periodic spontaneous structure at $A=0.6$, \eqref{eq:mu x}. The three modes are present above $T_c$. They correspond to the two instability modes, which are split due to explicit symmetry breaking (green and yellow), and a momentum relaxation mode, corresponding to the Drude peak (blue). The positions of the modes are obtained from the study of corresponding 2-point functions, which define the coloring. Below $T_c$ the leading instability triggers a phase transition and moves away, while the subleading instability plays a role of the damped Goldstone mode and mixes with the momentum relaxation mode, according to \cite{Delacretaz:2017zxd}. This mixing is evident from the fact that both modes appear in different correlators -- the blue and yellow dots overlap. Note the close similarity with time-dependent Ginzburg-Landau prediction, Fig.\,\ref{fig:U1QNMs}. At lower temperature the two imaginary modes collide and form a pair of real-valued modes, corresponding to propagating massive pseudo-Goldstone \cite{Andrade:2017cnc}. The darker markers represent the modes shown on Fig.\,\ref{fig:mode_profiles}. 
\\
\textbf{Right panel:}
The dependence of parameters on the amplitude of the ionic lattice, $A$, which corresponds to the scale of explicit symmetry breaking. For $A \lesssim 0.6$, the phase relaxation rate $\Omega$ (yellow) scales linearly $\Omega\,\sim\,A$, in agreement with \eqref{equ:def_modes_below_Tc}. The behavior of the momentum relaxation rate $\Gamma\,\sim\,A^2$ is also shown. The darker markers correspond to $A=0.6$, which is shown on the left panel. 
}
\end{figure}

The spectrum of the purely imaginary quasinormal modes of the example case at $A=0.6$ and $q_c=1.33 \mu$ is shown on Fig.~\ref{fig:periodic_modes}. See Appendix\,\ref{app:periodic} for details on obtaining this data. Note that the temperature range shown is extremely narrow: $\Delta T \sim 0.01 T_c$. In the normal phase one observes three purely imaginary modes. One of them (green) corresponds to the leading instability and ``bounces of'' the real axis at $T_c$, the other one (yellow) has the same slope above $T_c$ with a finite offset, being damped even at the point of phase transition. This is a subleading instability, which turns into a damped Goldstone mode below $T_c$. The third mode (blue) corresponds to a damped pole in the conductivity due to weak momentum relaxation and mediates the Drude peak above $T_c$. This spectrum is precisely as expected from the TDGL analysis of the previous Sec.\,\ref{equ:deformed_modes_above_Tc}, \eqref{equ:def_modes_below_Tc}, c.f. Fig.\,\ref{fig:U1QNMs}. 

\begin{figure}
\includegraphics[width=1 \linewidth]{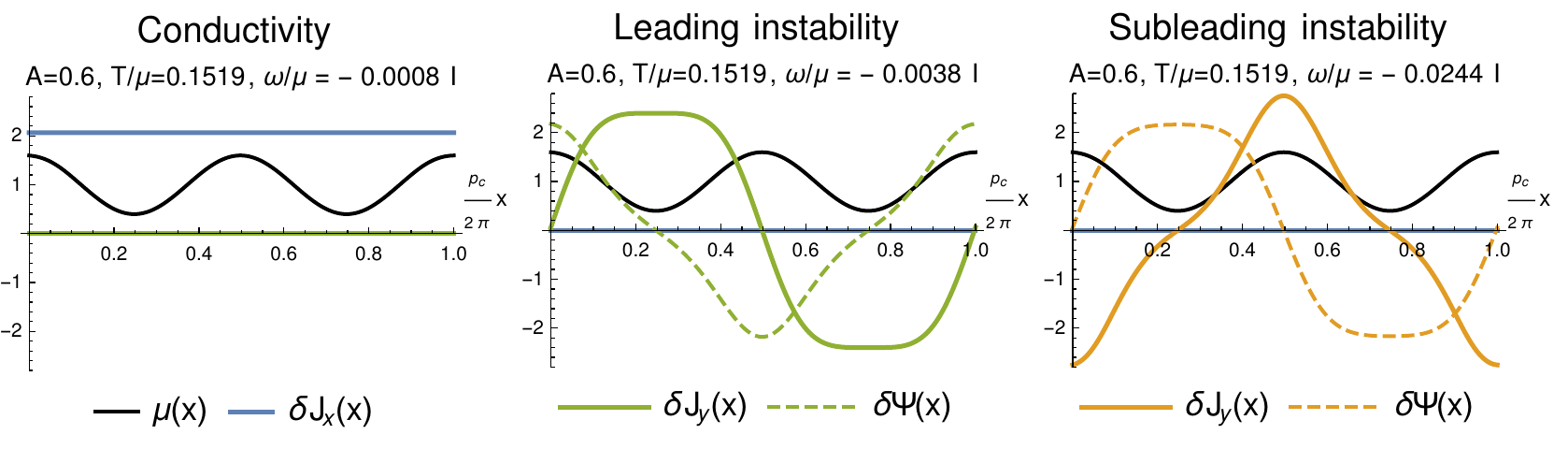}
\caption{\label{fig:mode_profiles}
\small
The profiles of the observables, corresponding to the quasinormal modes in the normal phase of holographic model with periodic spontaneous structure. The modes are those marked with darker markers on the left panel of Fig.\,\ref{fig:periodic_modes}. The black line shows the profile of explicit modulation of the background chemical potential \eqref{eq:mu x}. The conductivity mode (blue) is homogeneous and is sourced by a constant electric field $E_x$, giving rise to a constant current $J_x$. The two instability modes are triggered by the spatially oscillating sources $\delta \psi_1 \sim \cos(q_c x)$ (green), and $\delta \psi_2 \sim \sin(q_c x)$ (yellow). The observables are the transverse current $J_y$ (solid line) and the pseudoscalar operator $\Psi$, dual to field $\psi$ (dashed). Note the close similarity to the right panel of Fig.\,\ref{fig:toy_model_periodic}.}
\end{figure}

The identification of the modes is performed as follows: we study a set of two-point correlators, turning on different sources on the boundary, and figure out the imaginary frequencies where the correlators diverge. The momentum relaxation mode is triggered by the homogeneous source of the electric current $\mathcal{A}_x$, the leading instability mode is sourced by $\delta \psi_1 \sim \cos(p_c x)$, which is a part of \eqref{eq:marginal modes}, while the subleading instability is triggered by the source shifted by a quarter of a period $\delta \psi_2 \sim \sin(p_c x)$. The profiles of the observables in three modes are shown on Fig.\,\ref{fig:mode_profiles}. Even thought the effects of the background lattice induce the higher order harmonics in profiles of the instability modes, one can clearly identify them with the ones parametrized by real and imaginary parts of complex amplitude $A$ from our earlier model treatment \eqref{equ:ImReA}, see also right panel of Fig.\,\ref{fig:toy_model_periodic}. 

Below $T_c$ due to the appearance of a finite condensate, the momentum relaxation pole mixes with the damped Goldstone. Therefore in the broken phase both modes appear in the corresponding two-point functions, which were decoupled in the normal phase -- this effect is visible on Fig.\,\ref{fig:periodic_modes} when the yellow and blue dots overlap. At lower temperature the modes collide and form a pair of complex modes with finite real part -- the gapped massive pseudo-Goldstone. This behavior is well described by the effective theory of pinned charge density wave \cite{Delacretaz:2017zxd}. 

According to the TDGL treatment, the imaginary position of the damped Goldstone mode at critical temperature is directly proportional to the scale of the explicit symmetry breaking \eqref{equ:deformed_modes_above_Tc}. We check this prediction by tracing the QNMs for a range of amplitudes $A \in [0.2,1.2]$ (see additional data plots on Fig.\,\ref{fig:AllAQNMs} in Appendix\,\ref{app:periodic}). The result shown on Fig.\,\ref{fig:omega_A} confirms this prediction. The quadratic dependence of the momentum relaxation rate $\Gamma$ on the explicit TSB scale is also generally expected \cite{Hartnoll:2012rj}. 
Note also that the slopes of temperature dependence for the leading instability mode on Fig.\,\ref{fig:periodic_modes} agree with prediction of TDGL treatment \eqref{equ:def_modes_below_Tc}, \eqref{equ:deformed_modes_above_Tc}: below $T_c$ the slope is twice steeper then the one above $T_c$. 
Overall we conclude that the holographic model with periodic lattice follows the logic described in Sec.\,\ref{sec:model_TSB} quite precisely.

\section{\label{sec:MIT}Features of electric transport and metal-insulator phase transition}

\begin{figure}
\center
\includegraphics[width=0.49\linewidth]{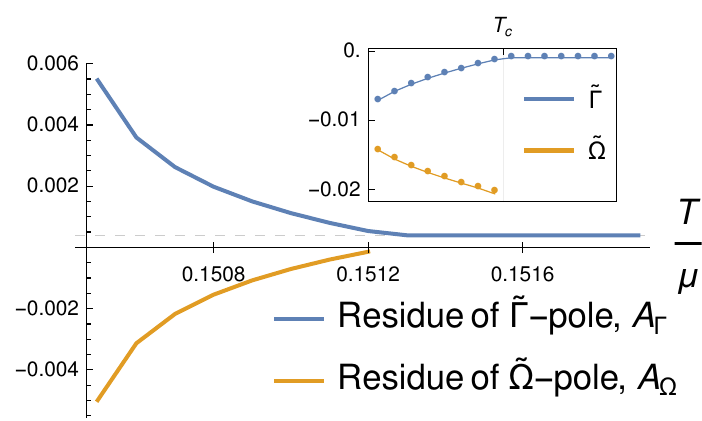}
\includegraphics[width=0.49\linewidth]{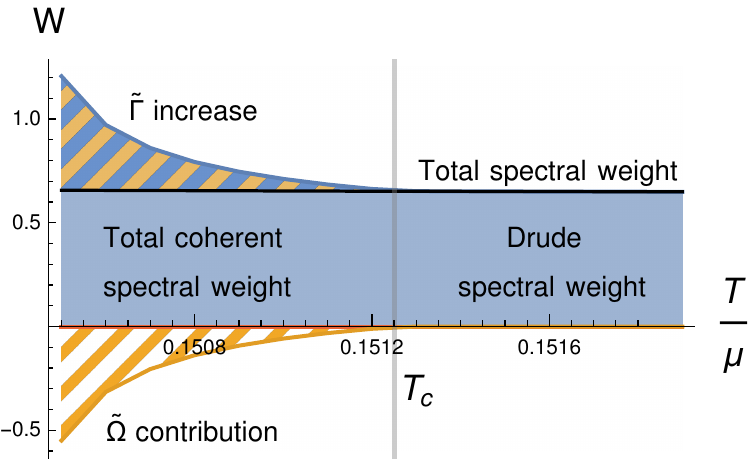}
\caption{\label{fig:spectral_weights}
\small
\textbf{Left panel:} The fitting parameters to the AC conductivity at real frequencies \eqref{equ:spectral_weights}. The positive residue of the $\tilde{\Gamma}$-mode, which corresponds to a Drude peak above $T_c$ gets significantly enhanced below $T_c$, while at the same time the $\tilde{\Omega}$-mode mixes in with negative residue, in agreement to \eqref{equ:model_residues}. The inset shows the excellent agreement between that the positions of the modes extracted from the fit (lines) and the ones obtained by the imaginary QNM analysis (dots), c.f. Fig.\,\ref{fig:periodic_modes}. The value of $\sigma_{\mathrm{inc}}$ is constant in this temperature range ($\sigma_{\mathrm{inc}} \approx 0.35$) and is not shown.
\\\textbf{Right panel:} Redistribution of the spectral weights across the phase transition. Despite the negative contribution of $\tilde{\Omega}$-mode (yellow hatched), it is exactly compensated by the increase of the $\tilde{\Gamma}$ weight (blue-yellow hatch), therefore the total coherent spectral weight (blue) is constant across $T_c$. Note also that the spectral weight due to incoherent part of conductivity (not shown) is insensitive to the phase transition.}
\end{figure}

\begin{figure}
\center
\raisebox{-0.5\height}{\includegraphics[width=0.49\linewidth]{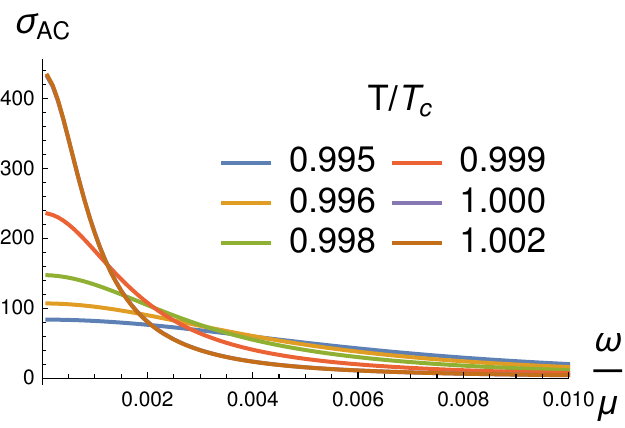}}
\raisebox{-0.52\height}{\includegraphics[width=0.49\linewidth]{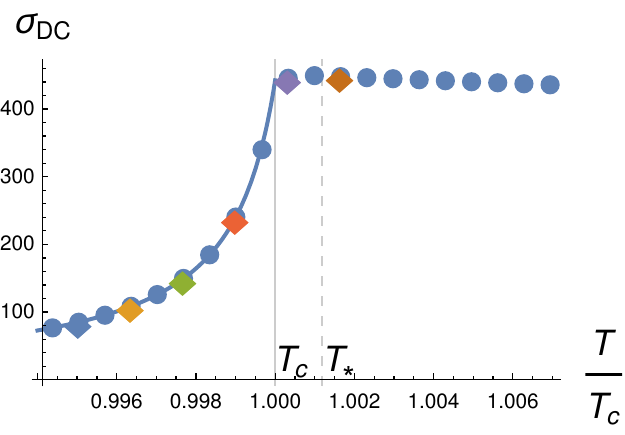}}
\caption{\label{fig:DC_around_Tc}
\small
\textbf{Left panel:} AC conductivity across the phase transition in periodic model. The sharp Drude peak above $T_c$ gets quickly deteriorated below $T_c$. \\
\textbf{Right panel:} DC conductivity across the phase transition. The blue dots correspond to the DC conductivity obtained directly from near horizon data \cite{Donos:2014cya}. The colored diamonds correspond to the zero frequency asymptotes of the AC conductivity curves from the left panel. Blue line is a fit with \eqref{equ:DC_at_TC}. The gridline shows a pole of the fit, which is located above $T_c$. A=0.6.}
\end{figure}

As we discussed above, a system with finite charge density, which develops spatially modulated charged structures in a broken phase, displays two features in the electric transport properties: one is governed by momentum relaxation and the corresponding Drude mode; and the other coming from the shifts of spontaneous structure -- the damped Goldstones which were in focus so far. We can now consider in more detail how exactly the two mechanisms contribute in the electric conductivity in a given holographic model. For this purpose we measure the AC conductivity of the system at a range of real frequencies for different values of temperature above and below $T_c$ (see Fig.\,\ref{fig:DC_around_Tc}). We observe that above~$T_c$ the AC conductivity is well fit by a single Lorentzian -- the Drude peak corresponding to the momentum relaxation rate $\Gamma$. Indeed, in the normal phase there is no spontaneous structure, and no charge is concentrated in it. Therefore the order parameter instability modes, which we studied in Sec.\,\ref{sec:periodic}, don't couple to the conductivity. Below $T_c$, however, the situation changes. Now the damped Goldstone mode starts overlapping with the current-current two point function and therefore the AC conductivity can be fitted as a sum of contributions from two imaginary poles:
\begin{equation}
\label{equ:spectral_weights}
\mathrm{Re}\,\sigma(\omega) = \sigma_{\mathrm{inc}} + \frac{A_\Gamma}{\tilde{\Gamma}^2 + \omega^2} + \frac{A_\Omega}{\tilde{\Omega}^2 + \omega^2},
\end{equation}
where $\sigma_{\mathrm{inc}}$ accounts for a constant background contribution which is not related to a coherent pole in the two-point function. It is ultimately related to the incoherent conductivity, mentioned in \eqref{AC model}, but for us now this is just a fitting parameter. Note also, that so far we don't assume any model interpretation for $\tilde{\Gamma},\tilde{\Omega}$ as well.
From the fit we observe a good agreement of the parameters $\tilde{\Gamma}$ and $\tilde{\Omega}$ with the positions of QNMs in the imaginary axis, which we studied above, see inset on the left panel of Fig.\,\ref{fig:spectral_weights}.
This measurement provides a good cross-check to our treatment and also shows that there are no other modes with finite real component, which we could miss in our scan for pure imaginary QNMs in case of periodic lattice, see Appendix \ref{app:periodic}. 

More importantly, with the fit of AC conductivity we can extract the contributions to electric transport from the two types of charge carrying processes: the movement of homogeneous charge density or shifts of charged spontaneous structure. The residues of the corresponding QNMs: $A_\Gamma$, $A_\Omega$ are shown in Fig.\,\ref{fig:spectral_weights}. As expected, the damped Goldstone mode only mixes into the current-current correlator in the broken phase, where the charge of the spontaneous structure is finite. The admixture happens smoothly as the residue of the $\tilde{\Omega}$ mode grows from zero at $T_c$ to a finite value. Interestingly, the contribution of this pole to conductivity turns out to be negative, therefore it can not be interpreted as arising from the finite density of extra independent charge carriers. Rather it points out that the two poles are substantially interacting with each other below $T_c$. This is not surprising given that indeed at slightly lower temperature they collide and form a pair of complex valued QNMs, corresponding to the pinned pseudo-Goldstone modes.

As we already mentioned in the Introduction, the effect of damped Goldstone on conductivity has been addressed in the framework of \cite{Delacretaz:2017zxd,Delacretaz:2016ivq} giving the expression with two poles below $T_c$ \eqref{AC model}.
The pinning frequency, $\omega_0$, provides the coupling between the two modes, which is proportional to the order parameter (and henceforth, the charge contained in the spontaneous structure) and the scale of explicit symmetry breaking \cite{Andrade:2017cnc}. 
It is the increase of $\omega_0$ below $T_c$, which drives the $\tilde{\Gamma}$ and $\tilde{\Omega}$ modes towards each other and ultimately introduces the gap in the spectrum of propagating pseudo-Goldstones at lower $T_c$, when the system develops finite elastic moduli.
Indeed, decomposing  Eq.\eqref{AC model} as a sum of poles \eqref{equ:spectral_weights},  we get in vicinity of $T_c$ ($\omega_0^2 \ll \Gamma \Omega$): 
\begin{equation}
	\tilde{\Gamma} = \Gamma + \frac{\omega_0^2}{\Omega - \Gamma} +  \dots, \qquad \tilde{\Omega} = \Omega - \frac{\omega_0^2}{\Omega - \Gamma}+ \dots.
\end{equation}
In accordance with the behavior which we observe on Fig.\,\ref{fig:spectral_weights}, the residue $A_{\tilde{\Omega}}$ is negative
\begin{align}
\label{equ:model_residues}
A_\Gamma &= \frac{\rho^2}{\chi_{\pi \pi}} + \frac{\frac{\rho^2}{\chi_{\pi \pi}} + 2 \gamma \rho (\Gamma - \Omega)}{(\Gamma-\Omega)^2} \omega_0^2 + \dots,&
A_\Omega &=
% - \frac{\frac{\rho^2}{\chi_{\pi \pi}} + 2 \gamma \rho (\Gamma - \Omega) + \chi_{\pi \pi} \gamma^2 (\Gamma-\Omega)^2}{(\Gamma-\Omega)^2} \omega_0^2 \equiv
- \frac{\left(\rho + \chi_{\pi \pi} \gamma (\Gamma - \Omega)\right)^2 }{\chi_{\pi \pi} (\Gamma-\Omega)^2} \omega_0^2 + \dots. 
\end{align}

From \eqref{equ:spectral_weights} one can obtain the total spectral weight in the conductivity channel $W=\int d \omega \mathrm{Re}[\sigma(\omega)]$. We do not include the contribution from $\sigma_{\mathrm{inc}}$ here, since it is constant in frequency and therefore should be subtracted. Moreover the incoherent conductivity is insensitive to the physics of translation symmetry breaking. Therefore it behaves continuously across $T_c$ and in a narrow window of temperatures is roughly constant. This is confirmed by our fitting results.
The spectral weight gets contributions from the two poles
\begin{align}
W_{\tilde{\Gamma}} &= \int \limits_0^\infty d \omega \frac{A_\Gamma}{\tilde{\Gamma}^2 + \omega^2} = \frac{\pi}{2} \frac{A_\Gamma}{\tilde{\Gamma}}, &
W_{\tilde{\Omega}} &= \int \limits_0^\infty d \omega \frac{A_\Omega}{\tilde{\Omega}^2 + \omega^2} = \frac{\pi}{2} \frac{A_\Omega}{\tilde{\Omega}}.
 \end{align}
Notably, despite the fact the $W_{\tilde{\Omega}}$ is negative, we see on Fig.\,\ref{fig:spectral_weights}, that it is balanced by the rise of $W_{\tilde{\Gamma}}$ and therefore the total spectral weight is always positive. Moreover it is manifestly constant across the phase transition\footnote{We thank Erik van Heumen for suggesting us to study this question.}. 

It is tempting to compare the behavior of this system to the usual treatment of pinned charge density wave in condensed matter \cite{RevModPhys.60.1129}. In that case after the Peirels instability occurs, the spectrum of the carriers acquires a gap. Therefore the DC conductivity is exponentially suppressed right below the critical temperature, signalling a very sharp metal-insulator phase transition. In the systems which we are considering, however, the behavior of conductivity in vicinity of $T_c$ (when $\omega_0^2 \ll \Gamma \Omega$) is governed by the the pinning frequency, which is proportional to the order parameter:
\begin{align}
\label{equ:DC_at_TC}
\sigma_{\mathrm{DC}} \bigg|_{T\rar T_c} &= 
%\sigma_{\mathrm{inc}}+\frac{\rho ^2}{\Gamma  \chi_{\pi\pi}}-\frac{\omega_0^2 (\Gamma  \gamma \chi_{\pi\pi}+\rho )^2}{\Gamma ^2 \chi_{\pi\pi} \Omega } 
\sigma_{\mathrm{inc}} + \frac{\rho^2}{\Gamma \chi_{\pi \pi}} \frac{1}{1 +  \omega_0^2 \frac{\left(1 + \Gamma \gamma \chi_{\pi \pi}/\rho \right)^2}{\Gamma \Omega}} + O(\omega_0^4), \qquad \omega_0^2 \sim (T_c -T).
\end{align}
We see that due to finite $\Omega$ the resistivity below $T_c$ drops algebraically as a function of $\omega_0^2$. This behavior is also confirmed by the results for DC conductivity which we obtain explicitly in holographic model, see Fig.\,\ref{fig:DC_around_Tc}. The sharp drop of the DC resistivity at the immediate vicinity of the phase transition is well fitted by a first order pole $(T-T_*)^{-1}$, where the formal parameter $T_*$ is larger then the critical temperature, in agreement to \eqref{equ:DC_at_TC}. This shape of the metal-insulator transition has already been observed earlier in \cite{Mott,Andrade:2017cnc,Andrade:2018gqk}.

The distinct behavior of the conductivity at metal-insulator transition, which we obtain, has profound origins. As we demonstrate in this work, the behavior of holographic models with pinned spontaneous spatial structures is well described by the time-dependent Ginzburg-Landau equation of Sec.\,\ref{sec:model_TSB}. While this approach is well motivated for classical systems and in many cases has been shown to emerge from the microscopic dynamics, it is also well known that TDGL is not appropriate for the quantum systems with gapped spectrum, see \cite{tinkham2004introduction,cyrot1973ginzburg} for a review. The reason is that the dynamics of the gapped system is dominated by the oscillation, rather then dissipation and therefore the dissipation equation \eqref{equ:time_GL} fails to describe it. This is the case in normal gapped superconductors \cite{tinkham2004introduction} and the conventional CDW \cite{RevModPhys.60.1129}, where the gap opens due to Peierls mechanism. In these systems the transport is indeed characterized by the gap and the conductivity is exponentially suppressed. However, as it has been shown for ``gapless superconductors'' in \cite{gor1996generalization}, as soon as the gap in the spectrum of elementary excitations is blurred out (by e.g. magnetic impurities), the TDGL can be rigorously derived from the microscopic Bardeen-Cooper-Schrieffer model. In other words, the fact that \eqref{equ:time_GL} is applicable in our case means that \textbf{the holographic charge density wave is ``gapless''}. This conclusion is in line with the earlier studies of holographic superconductors \cite{She:2011cm,Plantz:2015pem}, where the applicability of TDGL has also been addressed.

\section{\label{sec:helical}Holographic helical homogeneous lattice}

\begin{figure}
\center
\includegraphics[width=0.6 \linewidth]{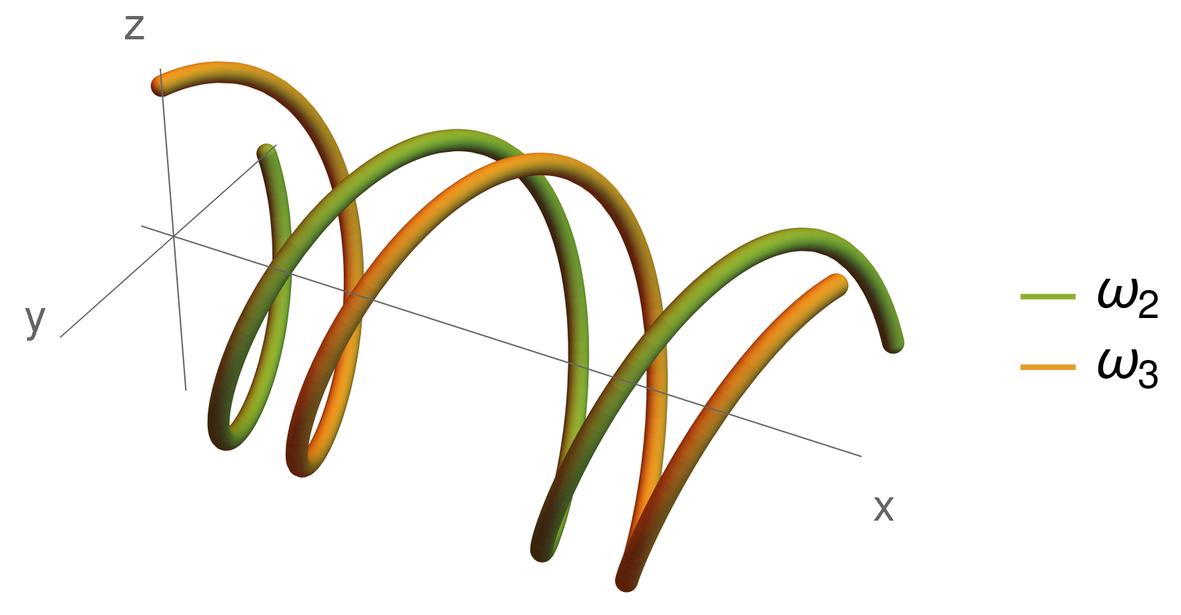}
\caption{\label{fig:helix}
\small
A cartoon of the helical structures $\omega^{(p_c)}_{2},\omega^{(p_c)}_{3}$, \eqref{equ:helical_forms}. The phase shift between them is manifest as a rigid translation in the $x$ direction.}
\end{figure}

Our conclusions derived above in the holographic model with periodic spontaneous structure are quite generic and in this Section we'll check whether they hold for the other example of holographic model with spontaneous dynamic TSB. Let's study the setup with helical Bianchy VII structure, which belongs to a class of homogeneous holographic models, mentioned in the Introduction, see also Sec.\,\ref{sec:homogeneous} below.
The model is defined in 5-dimensional bulk ($x^\mu = \{t,x,y,z,u\}$, $u$-- radial holographic coordinate, with boundary at $u=0$) with dynamical gravity, an Abelian gauge 
field $A_\mu$, dual to the chemical potential and an auxiliary vector field $B_\mu$, which can be used to source the explicit translational symmetry breaking. The action reads
\begin{equation}
\label{eq:action_helix}
  S = \int d^5 x \sqrt{- g} \left( R  - 2 \Lambda - \frac{1}{4} F^2 - \frac{1}{4} W^2  \right) - 
   \frac{\bm{\gamma}}{6} \int  A \wedge F \wedge F , 
   %- \frac{\kappa}{2} \int B \wedge F \wedge W,
\end{equation} 
where $\Lambda=-6$ and $F \equiv dA$, $W\equiv dB$ -- the field strength tensors. We set the mass of the $B$ field to zero and we chose $\bm{\gamma} = 3$ -- the same setup as in \cite{Andrade:2018gqk}. As it has been shown by \cite{Ooguri:2010kt, Donos:2012wi} the model in absence of explicit breaking source develops an instability at critical temperature, which leads to a formation of the structure characterized by the helical forms
\begin{align}
\label{equ:helical_forms}
\omega^{(p_c)}_1 & = dx \\
\notag
\omega^{(p_c)}_2 & = \cos (p_c x) dy - \sin(p_c x) dz \\
\notag
\omega^{(p_c)}_3 & = \sin (p_c x) dy + \cos(p_c x) dz. 
\end{align}
More precisely, in absence of explicit source, the two degenerate marginal modes appear which drive the instability:
\begin{align}
\label{equ:omega_modes}
\mbox{$\omega_2$ mode:} \qquad &\delta A_2 = \delta A_2(u)\, \omega_2^{(p_c)}, \qquad \delta(ds^2) = 2 \delta Q_2(u)\, dt\, \omega_2^{(p_c)},\\
\notag
\mbox{$\omega_3$ mode:} \qquad &\delta A_3 = \delta A_3(u)\, \omega_3^{(p_c)}, \qquad \delta(ds^2) = 2 \delta Q_3(u) \,dt\, \omega_3^{(p_c)}.
\end{align}
They describe the two patterns shifted by a quarter period, see Fig.\,\ref{fig:helix}, and any shifted helix with the same direction and pitch $p_c$ can be represented as a linear combination of these two linear modes, just like in \eqref{equ:ImReA}. Despite a little involved interpretation of the helical geometry, the Bianchy VII model has been widely used as a toy model for dynamical translation symmetry breaking thanks to the extra symmetries, which allow for its numerical treatment by means of ordinary differential equations (ODE) only \cite{Andrade:2015iyf,Andrade:2018gqk,Andrade:2017cnc}. 

When the explicit translation symmetry breaking is introduced via the source for the vector field $B_\mu$
\begin{equation}
\label{eq:bc B2}
B\big|_{u\rar0} = \lambda \omega^{(p_c)}_2,
\end{equation}
the system acquires finite resistivity in the normal state \cite{Donos:2012js, Donos:2014oha} and the degeneracy between the modes \eqref{equ:omega_modes} is lifted.  
It is quite important to note at this point, that according to \eqref{eq:action_helix} $B$-field only enters quadratically in the equations of motion for the spontaneously broken \hbox{$A$-field}. Therefore the corresponding explicit symmetry breaking term in the effective TDGL equation \eqref{equ:amplitude equation_deformed} is proportional to $f \sim \lambda^2$. Henceforth despite the fact that the explicit breaking is introduced via the same form $\omega^{(p_c)}_2$ as the spontaneous structure, the symmetry corresponding $\omega_2 \rar - \omega_2$ (equivalent to $\lambda \rar -\lambda$) is actually preserved. In this way we are effectively dealing with spontaneous $\mathbb{Z}_2$ residual symmetry breaking, similar to $2/1$ commensurate cases discussed above.

\begin{figure}
\center
\raisebox{-0.5\height}{\includegraphics[width=0.67 \linewidth]{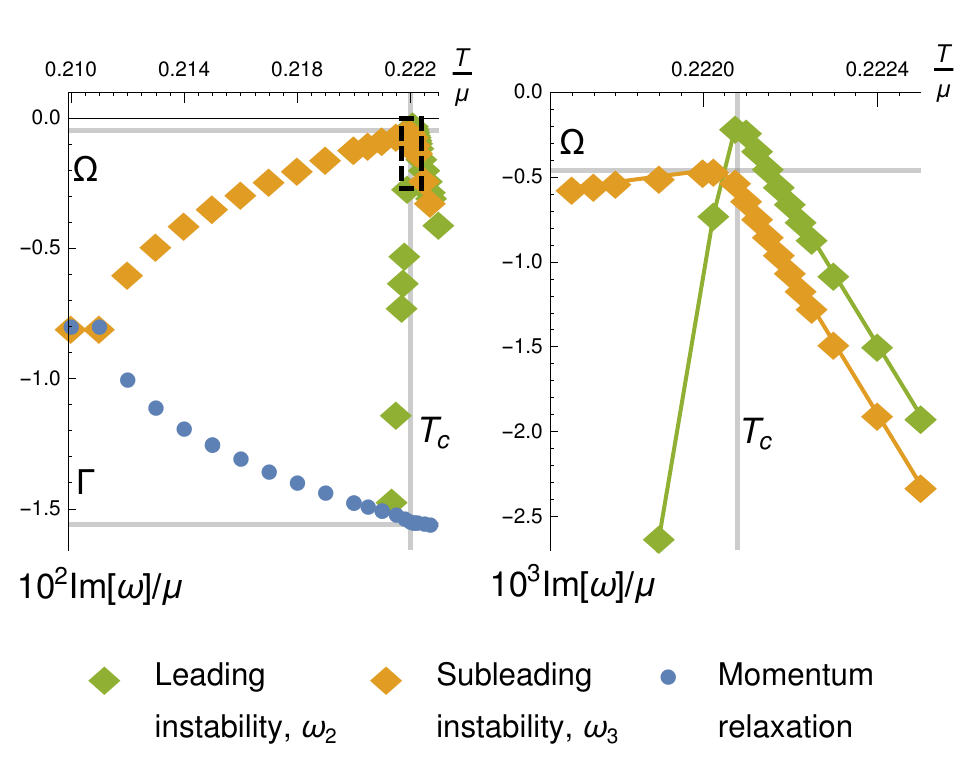}} \hfill
\raisebox{-0.54\height}{\includegraphics[width=0.32 \linewidth]{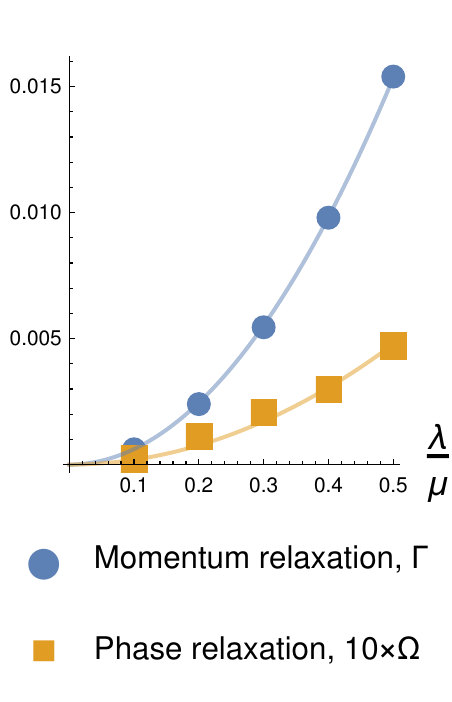}}
\caption{\label{fig:helix_spectrum}
\small
\textbf{Left panels:} The spectrum of purely imaginary quasinormal modes of the helical holographic model in vicinity of the critical temperature. The general behavior closely parallels that of Figs.\,\ref{fig:toy_model_periodic},\,\ref{fig:periodic_modes}. Three modes above $T_c$: momentum relaxation, leading instability, which drives the phase transition, and subleading instability, which turns into a damped Goldstone below $T_c$. The momentum relaxation and damped Goldstone modes become coupled below $T_c$ and recombine forming the pseudo-Goldstone at lower temperature. Note that, unlike Fig.\,\ref{fig:periodic_modes}, here $\Gamma \gg \Omega$. The data is for $\lambda=0.5 \mu, p_c = 2.16 \mu$. The second plot zooms into the dashed region of the first one. \\
\label{fig:helical_Omega}\textbf{Right panel:} Dependence of the phase relaxation rate $\Omega$ and momentum relaxation rate $\Gamma$ on the scale of explicit symmetry breaking $\lambda$ in helical model. Both behave quadratically.}
\end{figure}

The behavior of the quasinormal modes near critical temperature is shown on Fig.\,\ref{fig:helix_spectrum} for the model with $\lambda=0.5 \mu, p_c = 2.16 \mu$. Once again, we observe the familiar picture: the two split instability modes appear above $T_c$, which correspond exactly to $\omega_2$ and $\omega_3$ modes in \eqref{equ:omega_modes}. (We check this by the explicit analysis of their field profiles, see Appendix\,\ref{app:helix}.) One of them hits the real axis and triggers the phase transition to a broken phase, while the other freezes at finite imaginary value and becomes a damped Goldstone. In the broken phase this mode couples to electric current, due to a finite charge of the spontaneous helical structure and, as the temperature is lowered, recombines with the momentum relaxation mode, in order to form a massive pseudo-Goldstone. 
In case of helix our numerical method allows us to compute all QNMs including those with finite real part (see Appendix\,\ref{app:helix}), so we can follow this process. Moreover, we have explicitly verified that there are no other light modes in the complex plane apart from the described above. The non-hydrodynamic modes, which are generally present in the holographic computations, appear at $|\omega|\approx \mu$ and do not play any role in the physics under consideration.

Again, we can trace the dependence of the phase relaxation rate $\Omega$ on the scale of explicit TSB, $\lambda$, on the right panel of Fig.\,\ref{fig:helical_Omega}. It turns out that $\Omega \sim \lambda^2$, which might look surprising on the first sight, but actually agrees very well with the fact that the explicit symmetry breaking term appears as $f\sim \lambda^2$ in the equations of motion for the $A$-field -- as in the previous cases, we actually observe $\Omega \sim f$. The momentum relaxation rate~$\Gamma$, however, behaves in the usual quadratic way $\Gamma \sim \lambda^2$. 
Note that these results agree with our earlier studies conducted at low temperature \cite{Andrade:2018gqk}.

\begin{figure}
\center
\includegraphics[width=0.49 \linewidth]{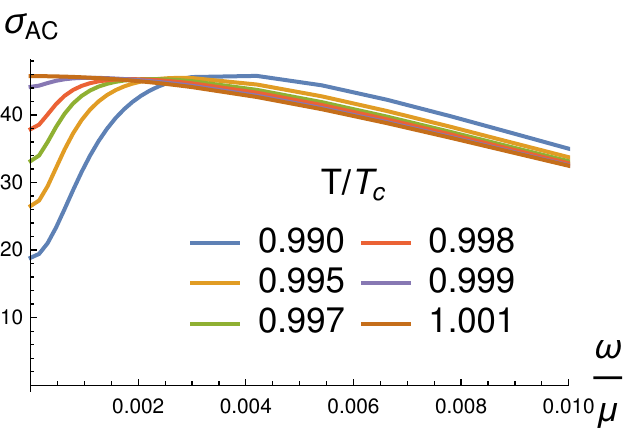}
\includegraphics[width=0.49 \linewidth]{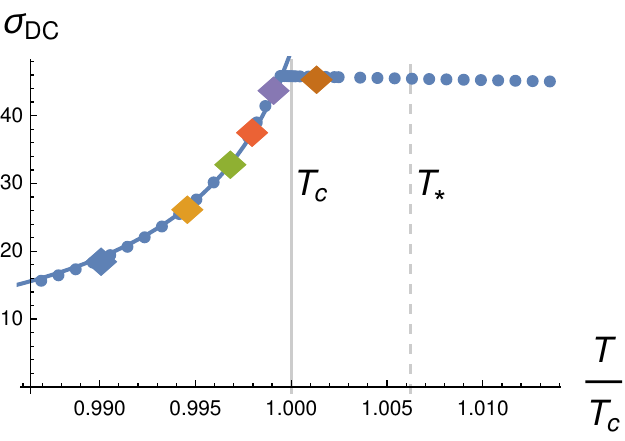}
\caption{\label{fig:helix_DC_at_Tc}\label{fig:helix_AC}
\small
\textbf{Left panel:} AC conductivity across the phase transition for the helical model. Unlike periodic case Fig.\,\ref{fig:DC_around_Tc}, a sharp depression develops in the relatively wide Drude peak below $T_c$. The loss of the spectral weight at small $\omega$ is compensated by the increase at larger $\omega$. We show the data for $\lambda=0.5 \mu, p_c =2.16 \mu$. $T_c \approx 0.2221 \mu$\\
\textbf{Right panel:} DC conductivity across the phase transition. The blue dots correspond to the DC conductivity obtained directly from near horizon data \cite{Donos:2014cya}. The colored diamonds correspond to the zero frequency asymptotes of the AC conductivity curves from the left panel. Blue line is a fit with \eqref{equ:DC_at_TC}. The gridline shows a pole of the fit, which is located above $T_c$. Note complete analogy with periodic case Fig.\,\ref{fig:DC_around_Tc}.  
}
\end{figure}

The DC conductivity in vicinity of critical temperature follows the logic described in Sec.\,\ref{sec:MIT} as well. Instead of a sharp exponential drop it demonstrates the algebraic behavior~\eqref{equ:DC_at_TC}, see Fig.\,\ref{fig:helix_DC_at_Tc}, right panel. Curiously, this drop is reflected in the profile of the AC conductivity (left panel of Fig.\,\ref{fig:helix_AC}) in a way which is quite different from the pattern we saw in Sec.\,\ref{sec:MIT}. Instead of a gradual degrease of the overall profile of the Drude peak, seen on Fig.\,\ref{fig:DC_around_Tc}, we observe a relatively sharp depression which develops in the AC conductivity profile at small $\omega$. This distinct behavior is due to the inverse hierarchy of $\Gamma$ and $\Omega$ scales in the helical model. While in periodic case we had $\Omega \gg \Gamma$ (see Fig.\,\ref{fig:periodic_modes}), in case of helix, on the contrary, we have $\Omega \ll \Gamma$, see Fig.\,\ref{fig:helix_spectrum}. Otherwise the behavior of the contributions from the two poles to the conductivity is exactly the same, which we can again figure out by fitting the AC curves with \eqref{equ:spectral_weights}. At $T>T_c$ only $\Gamma$ mode responsible for momentum relaxation contributes to $\sigma(\omega)$ leading to a relatively wide Drude peak. However at $T<T_c$ the extra contribution from $\tilde{\Omega}$ mode arises with negative residue \eqref{equ:model_residues}. This negative contribution is clearly visible as a sharp depression (with width of order $\tilde{\Omega}$), which develops at small frequencies. On Fig.\,\ref{fig:helix_AC} one can discern the rise of the $\tilde{\Gamma}$ residue, which results in a slight increase of the spectral weight at larger frequencies.

In the end of the day we see that the helical holographic model with dynamical translational symmetry breaking follows nicely the logic outlined by the TDGL treatment of periodic patterns in Section\,\ref{sec:model_TSB}. 
One might wonder how does it happen, since, as mentioned already, the helical model belongs to a class of holographic models with so-called ``homogeneous'' pattern of translation symmetry breaking and does not possesses true periodicity.
The homogeneous models contain operators which break translations but whose corresponding stress tensor remains coordinate independent. 
In this way, the background geometry is a function of the sole radial holographic coordinate and all the dynamics can be extracted by solving simple ODEs, as we saw here. 
The homogeneity and the lack of local structure in position space of these models generally result in the absence of well defined Brilluin zone in the spectrum of fluctuations and therefore the physics associated with umklapp scattering \cite{Bagrov:2016cnr}, or commensurate lock-in \cite{Andrade:2015iyf} can not be addressed. 

However, as we see it from the results for helical model, which are in perfect agreement with the treatment of Sec.\,\ref{sec:model_TSB}, the lack of of these features is irrelevant to the physics of the phase relaxation due to explicit symmetry breaking, which we address in this paper. 
In a way, this result is not completely surprising since
the microscopic crystal lattice structure (of length scale ``$a$'') is in principle irrelevant to the physics at long wavelengths and at larger scales any system would look homogeneous.
This is, for instance, the reason why the microscopic scale $a$ serves as the UV cutoff for any well-defined theory of phonons \cite{Leutwyler:1996er} or hydrodynamic theory for crystals and phases which spontaneously break translations \cite{chaikin2000principles,PhysRevA.6.2401}. The assumption of homogeneity is one of the basic principles of  the modern theory for solids built in \cite{Nicolis:2015sra}.
Given this fact, it is expected that both homogeneous helical model and the  inhomogeneous model of Sec.\,\ref{sec:periodic} give results in perfect agreement with the homogeneous hydrodynamic theories \cite{Delacretaz:2017zxd}.

\section{\label{sec:homogeneous}Holographic non-dynamic homogeneous lattices}

Let us now move forward and discuss other representatives of the family of homogeneous holographic lattices, the \textit{Q-lattice} and \textit{axions-like} models. In addition to absence of periodicity, these models may lack the mechanism for dynamical formation of the spatial structure, which was present in the helical case. However historically it is in these models, where the first observations of the physics related to the damped Goldstone mode has been made. Let us comment on the reason why our approach continues to hold in these simpler cases.

%Q-lattices
The Q-lattices were introduced in \cite{Donos:2013eha} and further studied in \cite{Amoretti:2017frz}. 
In these setups, a set of complex scalar fields, one for each spatial direction with broken translations, is introduced. The background solution is given by 
\begin{equation}
\label{equ:Q-lattice}
\Phi^I =\varphi(z)\,e^{i\,p_c\,x^I}\,,
\end{equation}
where $z$ is the holographic radial direction.
The action is invariant under a global $U(1)$ symmetry which simply shifts the phase of the scalars, therefore the coordinate dependence of the fields is lost in the stress-energy tensor, which guarantees the homogeneity of the geometry. Nonetheless, spacetime translations are broken by the profiles \eqref{equ:Q-lattice} 
and the corresponding physics can be adequately described \cite{Donos:2013eha,Donos:2014uba}. Moreover, the spontaneous TSB can also be addressed in the models where the dilaton field $\varphi(z)$ acquires a vacuum expectation value due to internal dynamics in absence of the source \cite{Amoretti:2017frz,Baggioli:2018bfa}. The presence of propagating phonons has been recently tested in \cite{Amoretti:2019cef}.
In presence of a small source for $\varphi$, the physics of pseudo-spontaneous TSB can be studied, including the softly gapped pseudo-Goldstone modes \cite{Amoretti:2018tzw,Amoretti:2019kuf,Donos:2019hpp}. Interestingly, since the slopes $p_c$ of the explicit and spontaneous structures are identical, the model can be though of as describing a fully commensurate state.

Note however, that unlike the helical model of Sec.\,\ref{sec:helical}, the wave vector $p_c$ in the \hbox{Q-lattice} models referred above is not chosen dynamically. It is rather a parameter of a particular ground state, which can be picked arbitrarily. Strictly speaking, the free energy is only minimized by the ground state with $p_c=0$, however since there is no dynamics associated with $p_c$, the states with finite wave vector are dynamically stable. In other words, in these models the spatial structure does not arise spontaneously. This disadvantage can be in principle circumvented by introducing higher derivative terms in the action \cite{Amoretti:2017frz,Baggioli:2014roa}, but one must be extra careful, since these modifications affect the dynamical stability of the ground state and can result in a negative shear modulus \cite{Ammon:2020xyv,Baggioli:2020ljz}.

Despite these deficiencies, the physics related to phase relaxation has been observed in the Q-lattice models even without dynamical wave vector. These homogeneous models display the presence of a diffusive Goldstone mode in the longitudinal sector whose origin is not related by any means to translational invariance. On the contrary, it has been shown explicitly that such mode comes from the spontaneous breaking of the global bulk symmetry \cite{Donos:2019txg}. 
In case when the soft explicit breaking is introduced, this hydrodynamic mode acquires a finite damping $\Omega$ which has been shown \cite{Amoretti:2018tzw} to depend crucially on the explicit breaking scale and therefore can be thought as our familiar \textit{damped phason} mode. 
More precisely, it has been shown that its relaxation rate obeys a universal correlation:
\begin{equation}
\Omega\,\sim\,\chi_{\pi\pi}\,\omega_0^2\,\Xi\,, \label{cc}
\end{equation}
involving the phonons pinning frequency $\omega_0$ and the parameter $\Xi$ which governs the diffusion constant of the phason in absence of explicit symmetry breaking. This relation has been checked in several models and it is indeed very robust \cite{Amoretti:2018tzw,Ammon:2019wci,Donos:2019hpp,Amoretti:2019cef,Andrade:2018gqk}. A partial derivation for it was presented in \cite{Baggioli:2020nay} and a final proof in \cite{Baggioli:2020haa}. Interestingly, the implicit observation in the context of incommensurate crystals already appeared in \cite{Currat2002}. Note that \eqref{cc} is valid for helical model considered above as well: there, what concerns the dependence on the explicit TSB scale, $\Omega \sim \lambda^2$ and $\omega_0 \sim \lambda$ \cite{Andrade:2017cnc}, while $\Xi$ and $\chi_{\pi\pi}$ don't depend on $\lambda$, since these features are not related to explicit TSB.

The relation \eqref{cc} follows directly from the lock-in of the spacetime translations and the internal global symmetry group. Indeed, we can understand the origin of the Q-lattice $\Omega$ mode in terms of Sec.\,\ref{sec:model_TSB} by taking into account the extra $U(1)$ symmetry of the model. 
%In the Q-lattice models, the global symmetry is a U(1) symmetry of the complex scalar fields. 
Indeed, the shift in $x^I$ can be canceled by a constant phase rotation of $\Phi^I$. (In case of helix the same role is played by the spatial rotations in $(y,z)$-plane). In particular, the background solution \eqref{equ:Q-lattice} can be parametrized with a complex-valued amplitude $\Phi_0$:
\begin{equation}
\Phi = \Phi_0 e^{i p_c x} \equiv  (\phi_1 + i \phi_2) e^{i p_c x}.
\end{equation}
In complete analogy with the pattern formation treatment of Sec.\,\ref{sec:model_TSB}, the shifts in \hbox{$x$-coordinate} are equivalent to the shifts of complex phase of $\Phi_0$
\begin{equation}
\label{equ:Q_lattice_shifts}
x \rar x + \delta x \quad \Leftrightarrow \quad \Phi_0 \rar \Phi_0 e^{i \delta \vartheta}, \qquad \delta \vartheta = \delta x/ p_c.
\end{equation}
Once again, when the ground state \eqref{equ:Q-lattice} is formed below $T_c$ with real $\Phi_0 = \varphi(z)$, the fluctuations of phase $\delta \vartheta$ are accounted for by the imaginary component $\delta \phi_2$, which acquires a finite imaginary part in presence of explicit source, see also discussion in Appendix\,\ref{app:TDGL}.

Let us compare this situation with the setup discussed in Sec.\,\ref{sec:model_TSB}. When the spontaneous and explicit breaking are driven by the same bulk field, in the normal phase there is no residual $\mathbb{Z}_2$ discrete symmetry, which would correspond to flipping the sign of the amplitude\footnote{Recently, more complicated models with additional sectors have been proposed to exactly disentangle the spontaneous and explicit dynamics \cite{Amoretti:2019kuf,Donos:2019hpp}. There, we do expect the $\mathbb{Z}_2$ symmetry to be present in total analogy to our helical models.}. 
This symmetry broken by the explicit source and it does not break spontaneously at the critical temperature. In the language of \cite{PhysRevLett.56.724}, this seems to correspond to a 1/1 commensurate case with no residual symmetry. As we discuss in more detail in  Appendix \ref{app:TDGL}, following the results of \cite{PhysRevLett.56.724}, we can verify that the difference between the 1/1 case and the 2/1 case is indeed the absence of a well-defined phase transition, but the dynamics of the modes and in particular of the phase relaxation $\Omega$ is qualitatively similar. This is the reason why the results outlined above for the Q-lattices  fit well in our framework. Note also, that our TDGL treatment does not assume any dynamical nature of the wavenumber of the spontaneous structure in \eqref{equ:ImReA}, so it is indeed totally applicable to the case of Q-lattice and explains the origin of $\Omega$-mode there.

Finally, we can address the most minimalistic holographic lattices, \textbf{the axions-like models} introduced in \cite{Andrade:2013gsa} and later generalized in \cite{Baggioli:2014roa,Alberte:2015isw}\footnote{See also \cite{Grozdanov:2018ewh} for a dualized version of this model which makes use of higher-forms symmetries.}. 
Axion-like theories have very strong similarities with the effective field theory of \cite{Nicolis:2015sra} and are constructed using a set of massless real scalars $\phi^I$ with a linear in the (spatial) coordinates profile
\begin{equation}
\label{equ:axions}
 \phi^I=p_c x^I. 
\end{equation}
The models enjoy a global internal shift symmetry $\phi^I\rightarrow \phi^I+\delta \vartheta^I$ (c.f. $U(1)$ in Q-lattice and $(y,z)$-rotations in helix) which is broken together with spacetime translations $x^I\rightarrow x^I+\delta x^I$ to their diagonal subgroup $\delta \vartheta^I=-\delta x^I$, c.f. \eqref{equ:Q_lattice_shifts}. The shift symmetry is guaranteed by constructing the action only with derivatives of those fields. The model exhibits the presence of propagating phonon modes \cite{Alberte:2017oqx} with speeds in agreement with the theories of elasticity \cite{chaikin2000principles} and hydrodynamics \cite{PhysRevA.6.2401}.

Unlike all the models considered above, the axion model does not have any kind of dynamical phase transition. Not only the critical momentum $p_c$ is not selected dynamically, but even the critical temperature is absent. In other words, these models only describe the broken phase. Nonetheless, and here it is the beauty of these models, all the physics previously described below the critical temperature $T_c$ is appearing and is in total agreement with the other more complicated models \cite{Alberte:2017cch,Alberte:2017oqx,Ammon:2019wci,Ammon:2019apj}. 
Since the dynamical phase transition is absent, the amplitude mode is not present in the spectrum, and only the Goldstone mode related to $\delta \vartheta$ remains. One can, again, decompose the scalar profile in the broken phase as
\begin{equation}
\phi^I = \phi_0^I + p_c x^I,
\end{equation}
With $\phi_0^I = 0$ in the ground state. The phase shifts, corresponding to a global shift symmetry, would result in $\phi_0^I \rar 0 + \delta \vartheta^I$. The corresponding hydrodynamic mode is a diffusive phason mode which has been observed directly in the QNMs spectrum \cite{Ammon:2019apj,Baggioli:2019abx,Baggioli:2020edn} and which played a major role \cite{Ammon:2020xyv} in identifying the correct hydrodynamic description of these gravitational setups \cite{Armas:2019sbe}. In presence of an explicit source, which in these models corresponds to deforming the axions-potential, this diffusive mode becomes damped with a relaxation rate given by $\Omega$. Interestingly, even in these simple models, it follows the universal relation \eqref{cc}.
Given the presence of the diffusive phason in their spectrum, it is not surprising that these simple models are perfectly describing the physics of phase relaxation below $T_c$, while they are missing the dynamics of the instability which is indeed driven by the other (and here absent) $\delta \phi_1$ amplitude mode. Moreover, as in the standard Q-lattice models, here there is no residual $\mathbb{Z}_2$ symmetry and the system can be thought to be in a 1/1 commensurate phase.

Finally, let us draw some general lessons. (I) The dynamics in the hydrodynamic regime (low frequency/momentum) is completely insensitive to the inhomogeneous nature of the system. As such, homogeneous and non-homogeneous models share exactly the same features. (II) The physics of phase relaxation does not rely on the presence of a well-defined phase transition nor of a dynamically chosen preferred wave-vector.
 (III) 
 The features of transport in phases which break translations spontaneously and/or explicitly in homogeneous models are not an artifact of the homogeneity assumption. However, in order to fully flesh out their nature, it is useful to study them in the more general context of inhomogeneous models.
 %} 
%

\section{\label{sec:conclusion}Conclusion}

In this work, we investigated the nature of the damped Goldstone (phason) in the spectrum of holographic models with spontaneous translation symmetry breaking and the associated phase relaxation rate $\Omega$, which has previously been observed in several homogeneous holographic models with broken translations. Our scope was to understand the meaning of this feature and its origin across the dynamical phase transition. 
We found out that the damped Goldstone can be well understood in terms of the general framework of  the time dependent Ginzburg-Landau equation. The two holographic models -- with periodic and helical patterns -- which we have considered in detail, show remarkable agreement of their quasinormal mode spectra with the one predicted by the TDGL.

We see that the phase relaxation mechanism is robust and persists both above and below critical temperature.
Its origin can be confidently attributed to the fluctuations of the two components of the spontaneous order parameter, which get split in the presence of a small explicit breaking source.
While one of the modes is driving the instability, the other plays a role of the damped phason in the broken phase. The physics of this second mode is continuous along the phase transition and couples to the electric transport below $T_c$ making the $\Omega$ pole to appear in the denominator of the AC conductivity formula \eqref{AC model}.
We also show how this new dynamics is responsible of the algebraic suppression of the DC conductivity at the metal-insulator transition, which appears in contrast to the standard exponential suppression of the conductivity for pinned charge density waves \cite{RevModPhys.60.1129}.

Two important observations follow from our study. Firstly, we show that the amplitude equation \eqref{equ:amplitude equation_deformed}, the version of TDGL equation applied for classical pattern forming systems \cite{PhysRevLett.56.724}, describes very well the dynamics of the holographic models with spontaneous spatial structures. At first sight this does not seem surprising, since the TDGL approach is based on the reasonable assumption that the system always tends to its thermodynamic equilibrium.
Moreover, the validity of TDGL-like dynamics has long been known in the field of classical gravity under the name of Gubser-Mitra conjecture \cite{Gubser:2000ec} for black branes. The conjecture states that thermodynamic equilibrium defined by a minimum of the free energy \eqref{equ:GL} implies the dynamical stability defined as the zero time derivative in \eqref{equ:time_GL}. 
In other words (see \cite{Emparan:2012be} for a more recent perspective), the conjecture holds if the modes that trigger the instabilities are long wavelenght modes, which are captured by the hydrodynamic analysis% 
\footnote{Counterexamples correspond precisely to gapped unstable modes, which drive dynamical instabilities and are transparent to the thermodynamics. Note also that 4-derivative model of \cite{Amoretti:2017frz,Baggioli:2014roa} seems to violate the conjecture given its instability \cite{Ammon:2020xyv,Baggioli:2020ljz}.}.
As we have argued, the transport properties of the systems of interest to us are well-described in the hydro approximation, which in turn justifies the applicability of TDGL. 

However from the point of view of the quantum system, which is dual to our gravitational model, the validity of TDGL equations below the critical temperature is quite exceptional. As we discussed it already, the conventional condensed matter systems with CDW develop a gap in the spectrum of elementary excitations \cite{RevModPhys.60.1129}. This gap on one hand makes the system an insulator with exponentially suppressed conductivity, while on the other hand it makes it impossible to derive TDGL-like dynamics from the microscopic theory: the system doesn't have excitations with arbitrary small energy, which would mediate its approach to equilibrium. The dynamics of the gapped quantum system is oscillating, and not diffusive. The situation, when TDGL can be derived in quantum system in the broken phase, arises when the gap is smeared out by impurities -- the ``gapless superconductor'', discussed in \cite{gor1996generalization}. So in a sense the holographic models, which we discuss present a ``gapless'' charge density wave. 
This statement is not particularly novel in holography. Indeed it has long been known that in holographic models the very existence of stable quasiparticle is challenged, therefore speaking about its spectrum and the associated gap is inadequate. Moreover, the holographic models generically possess the nonzero incoherent spectral density, the ``quantum critical continuum'', which allows for the relaxation processes, which would be forbidden in a more conventional quantum systems. For instance, it has been demonstrated in \cite{Krikun:2018agd,Baggioli:2019aqf} that due to this extra spectral density the, otherwise protected, plasmon excitation acquires anomalous width. Spectacularly, here we observe how this generic feature of holographic models affects the physics of pinned charge density waves.

The other important observation which we make in this work is a striking similarity between the behavior of various holographic models, starting from periodic inhomogeneous with dynamical translation symmetry breaking all the way to homogeneous linear axions, as predicted by the hydrodynamic theory for pinned incommensurate CDW \cite{Delacretaz:2017zxd}.
Using the series of holographic models we managed to isolate the physics of phase relaxation, 
which in the end of the day is insensitive to the local structure of the symmetry breaking pattern or the dynamics of phase transition.
This confirms the idea that the physics at long wavelengths is always homogeneous as originally claimed in \cite{Nicolis:2015sra}. The key feature, which is present in all the models we studied, is a certain global symmetry which is broken explicitly by external sources and spontaneously by a ground state. 
In fact, any spontaneous structure possesses a global shift symmetry corresponding to rigidly sliding it on top of the explicit structure. Even in the inhomogeneous model considered in this paper there is a global symmetry which corresponds to sliding the spontaneous structure on top of the ionic lattice.
The same pattern appears in the hydrodynamic formulations \cite{Delacretaz:2017zxd,Armas:2019sbe}.
This global symmetry is the ultimate responsible for the phase fluctuations mode and the $\Omega$ relaxation term. The physics of this diffusive mode can be related to that of quasicrystals \cite{Baggioli:2020nay} -- aperiodic structures -- and it can be thought to parametrize the diffusive phason mode in incommensurate structures. 
An EFT formulation explaining its diffusive nature at long wavelength has been recently constructed using Keldysh-Schwinger techniques \cite{Baggioli:2020haa} and discussed in \cite{Landry:2020ire}.

In summary, our work combines standard techniques borrowed from pattern formation theory with novel holographic methods and it provides a final understanding of the damped $\Omega$ phason mode. As already stressed, this dynamics is not a simple ''holographic copy'' of known condensed matter mechanism, but it introduces a totally new set of phenomena, whose experimental consequences have yet to be disclosed.

A few questions remain open, which deserve further investigation. As we see, the dissipative behavior of our system can be attributed to the effect of holographic quantum critical continuum, which smears the gap and allows for relaxation. In case of translational symmetry breaking, the order parameter is characterized by a finite wave-number $p_c$, so its relaxation dynamics actually probes the spectral density at finite momentum. It would be interesting to compute this spectral density in holographic model directly, in order to square up the above argument.

Another question concerns the fundamental role of global symmetry in holographic homogeneous models. Generically this global symmetry is expected to be a symmetry of the dual field theory. Following the holographic dictionary, it should be gauged in the bulk, i.e. associated to some conserved currents. This is what is done in \cite{Esposito:2017qpj}. Nevertheless, all the homogeneous models keep it as a global symmetry in the bulk, whose dual interpretation is not totally clear. One interpretation is that such global symmetry corresponds in the dual field theory to a symmetry with no associated Noether current \cite{Baggioli:2020haa}, but further investigation is desirable.

It would be also quite interesting to investigate how our treatment is modified when several such global symmetries occur in a situation when the spontaneous ordering happens in more then one direction, like in holographic minimally packed phases \cite{Donos:2015eew}, checkerboards \cite{Withers:2014sja}, or Abrikosov lattices \cite{Donos:2020viz}. In principle the 2-dimensional patterns display much richer structure and the interplay between the different shift symmetries could be quite nontrivial.

Finally, one could further investigate the role of commensurability and dynamics of the phason mode across an commensurate-incommensurate transition, discussed in \cite{Mott,Krikun:2017cyw}. In the present work, we have unnaturally fine-tuned the wave-vectors of the spontaneous and explicit structures to be equal; it would be interesting to follow the complete dynamics by allowing an initial incommensurate configuration. We plan to revisit this issue in the near future.

\acknowledgments{
We thank Daniel Arean, Blaise Gout\'eraux, Sebastian Grieninger, Michael Landry, Li Li, Jan Zaanen, Koenraad Schalm, Floris Balm and Erik van Heumen for enlightening discussions.
T.A. is supported by the ERC Advanced Grant GravBHs-692951.
A.K. is supported by VR Starting Grant 2018-04542 of Swedish Research Council and in part by RFBR grant 19-02-00214 of Russian Foundation for Basic Research.
A.K. appreciates the opportunity to deliver and discuss the preliminary results of this study, as part of the HoloTube seminar series.    
A.K. acknowledges the hospitality of Institute of Theoretical Physics in Autonomous University of Madrid where this project has been initiated. 
M.B. acknowledges the support of the Spanish MINECO ``Centro de Excelencia Severo Ochoa'' Programme under grant SEV-2012-0249. 
M.B. thanks the organizers of ``Quantum Matter and Quantum Information with Holography'', where part of this material was discussed, for the invitation.
Various parts of the numerical computations were performed on the resources provided by the Swedish National Infrastructure for Computing (SNIC) at SNIC Science Cloud partially funded by the Swedish Research Council through grant agreement no. 2016-07213 and the Maris Cluster of Instituut Lorentz, Leiden University.
}

\appendix

\section{\label{app:periodic}Inhomogeneous periodic lattice model}

\begin{figure}
\center
\includegraphics[width=1 \linewidth]{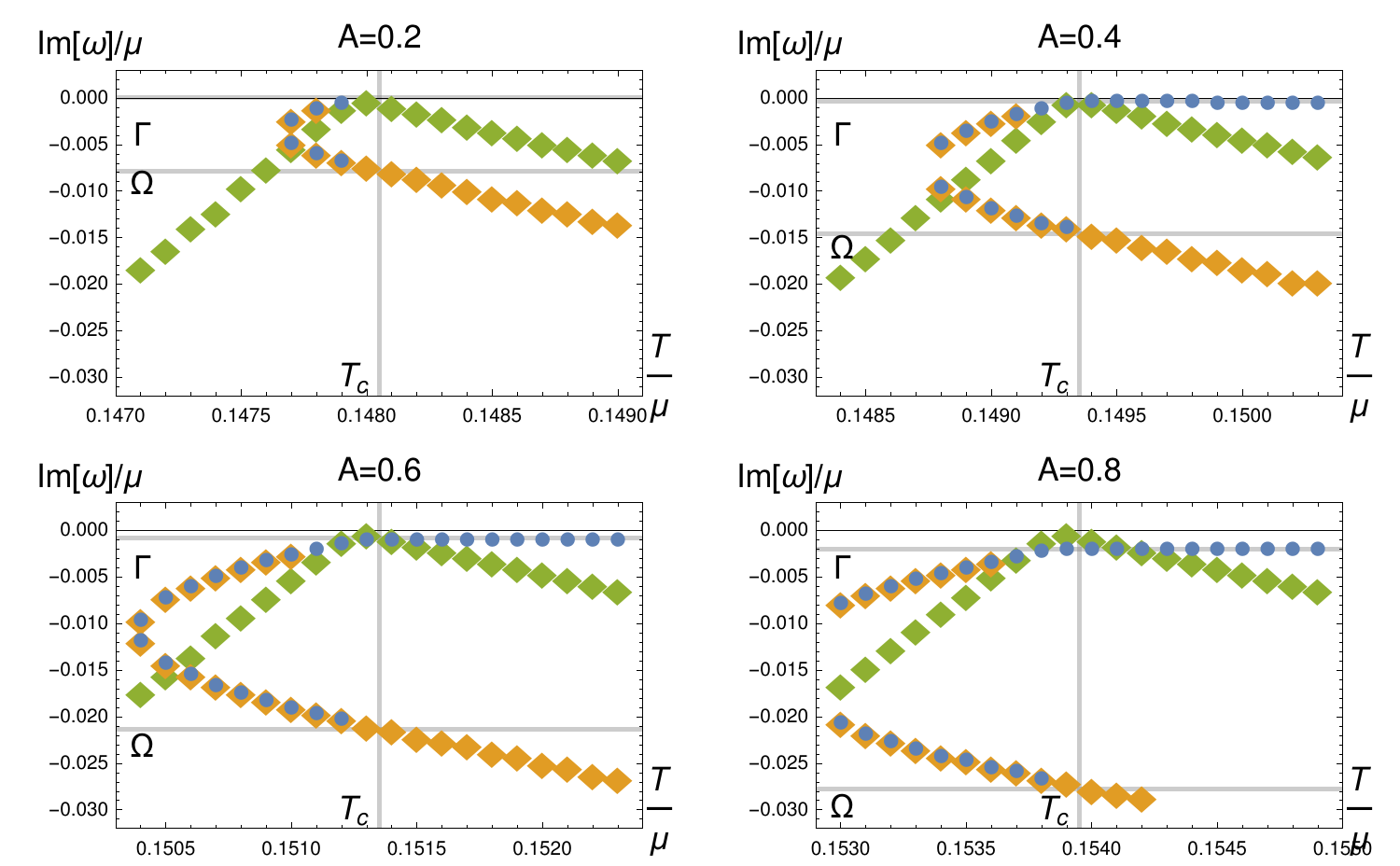}
\caption{\label{fig:AllAQNMs} 
\small
The results for the spectrum of quasinormal modes, obtained for different periodic lattice amplitude. The evolution of the spectrum is clearly visible. The data for $\Omega$ and $\Gamma$ are used to plot the corresponding dependencies on the right panel of Fig.\,\ref{fig:omega_A}}.
\end{figure}

In this work we use the holographic model with spontaneous periodic structure and ionic lattice, discussed in \cite{Mott,Krikun:2017cyw} with the same parameters, so much of the details regarding the ground states can be found in those works. In absence of the explicit periodic potential the phase transition at $T^0_c$ is driven by marginal mode with momentum $p_c$
\begin{equation}
p_c = 1.33 \mu,
\end{equation}
precisely in the same way as it is discussed in classical pattern formation literature \cite{cross1993pattern, PhysRevLett.56.724}. On top of the spontaneous structure we introduce the explicit external potential, which has a period aligned with the charge modulations in the spontaneous structure, i.e.
\begin{equation}
k_0 = 2.67 \mu.
\end{equation}
This corresponds exactly to a case of $2/1$ commensurability, discussed in \cite{PhysRevLett.56.724} and Sec.\,\ref{sec:model_TSB}. 
In this work we consider the background lattices \eqref{eq:mu x} with the amplitudes $A=\{0.2, 0.4, 0.6,0.8\}$ and study the solutions at temperatures in the immediate vicinity of the phase transition, see extra data on Fig.\,\ref{fig:AllAQNMs}. Note that the critical temperature is affected by the external potential (see discussion on $T^f_c$ vs $T^0_c$ in Sec.\,\ref{sec:model_TSB}) and we have $T_c = \{0.181,0.1494,0.1514,0.1540\}$ correspondingly.

In order to study the quasinormal modes and conductivity in the system we consider the linear perturbations of the fields on top of the background solution. Assuming the DeDonder gauge \cite{Rangamani:2015hka}, we include the time dependent perturbations of 10 components of the metric $\{\delta g_{tt},\delta g_{xx},\delta g_{yy},\delta g_{zz},\delta g_{tx},\delta g_{ty},\delta g_{tz},\delta g_{xy},\delta g_{xz},\delta g_{yz} \}$, 4 components of the gauge field $\{\delta \mathcal{A}_t, \delta \mathcal{A}_x, \delta \mathcal{A}_y, \delta \mathcal{A}_z \}$ and a perturbation of the pseudoscalar field $\delta \psi$ -- in total 15 modes, each of which is a function of holographic $z$ and spatial $x$ coordinates. Therefore the problem of the fluctuation analysis boils down to solving a system of 15 linear partial differential equations. 

\begin{figure}
\center
\includegraphics[width=0.49 \linewidth]{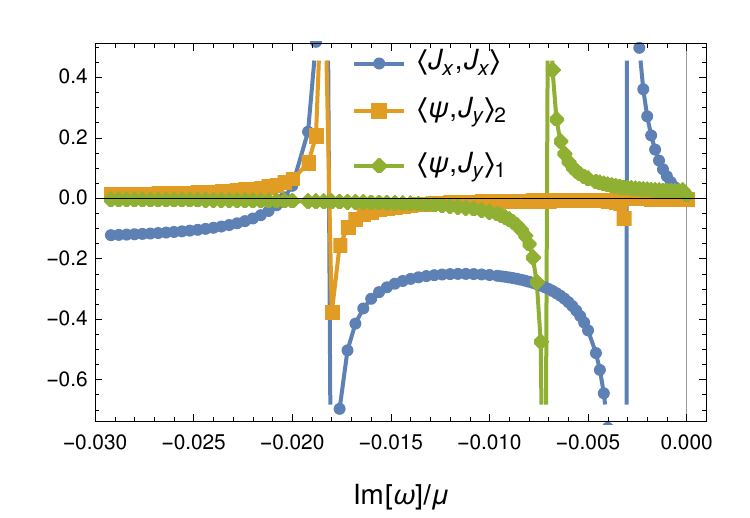}
\includegraphics[width=0.49 \linewidth]{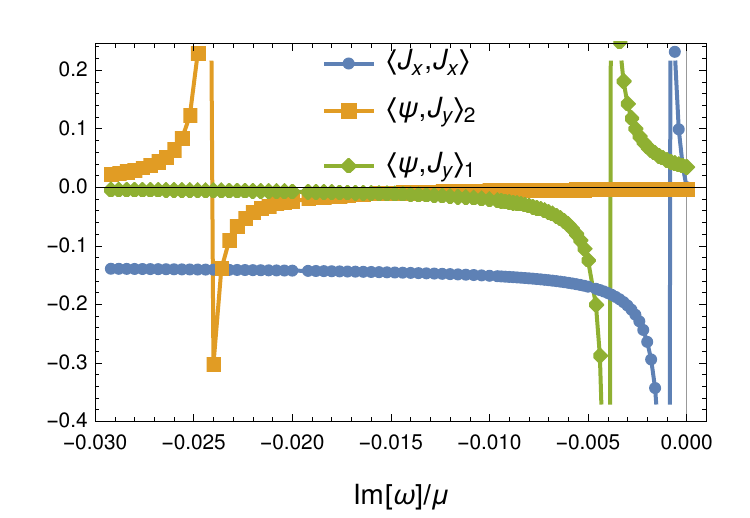}
\caption{\label{fig:imSigmaScan} 
\small
The profiles of the 2-point correlation function at imaginary frequency, used to detect the positions of quasinormal modes shown on Fig.\,\ref{fig:periodic_modes}. 
\textbf{Right panel:} Above $T_c$ the modes are decoupled and every single peak appears in its own channel. 
\textbf{Left panel:} Below $T_c$ the conductivity channel, dominated by the momentum relaxation mode, overlaps with shifts of spontaneous structure described by the damped Goldstone mode, therefore the corresponding QNMs appear in both channels. The data is for $A=0.6$, the temperatures are $T/\mu = 0.1509$ and $T/\mu = 0.1519$, respectively.
}
\end{figure}

In principle, one could obtain all the quasinormal modes of the system by evaluating the generalized eigenvalues of the matrix characterizing the fluctuation equations discretized on a computation grid. We use this method in case of helical model, where the equations are ODEs. However for PDEs this task turns out to be too numerically demanding, so we adopt a different method instead. In order to find the purely diffusive quasinormal modes it is enough to evaluate the corresponding two point function at the imaginary axis of complex frequency. Given that the QNMs result in the poles of the two point function, we can directly observe them in the imaginary frequency scan, see Fig.\,\ref{fig:imSigmaScan}.

We focus on the 3 different correlation functions. $\la J_x J_x \ra$ -- is the current-current correlator, which accounts for electric conductivity. The other two correlators are related to a finite wave-vector order parameter susceptibility. As we discussed in the main text, the marginal mode, which drives the instability and the order parameter itself include mainly 3 fields: $\delta \psi$, $\delta \mathcal{A}_y$ and $\delta g_{ty}$. The fields $\delta \psi$ and $\delta \mathcal{A}_y$ interact linearly in the linearized fluctuation equations, therefore the most direct way of measuring the quasinormal modes in the order parameter susceptibility channel is to take the cross-correlation $\la \Psi J_y \ra$ (where the operator $\Psi$ is dual to bulk pseudoscalar $\psi$ and the current $J_y$ is dual to $\mathcal{A}_y$). At finite wave-vector, however, it is useful to decompose the position dependent perturbative sources for the field $\delta \psi$ into $\sin$ and $\cos$ modes, in spirit of Sec.\,\ref{sec:model_TSB}. So in the end of the day in order to obtain the 3 correlation functions we turn on the sources
\begin{equation}
\label{equ:sources}
\delta A_x (x) \Big|_{z = 0} = J_x, \qquad ( z \delta \psi (x))\Big|_{z = 0}= \Psi_{1} \cos(p_c x) + \Psi_{2} \sin(p_c x),
\end{equation}
and measure the corresponding responses from the solutions of the linearized equations of motion. The profiles for the 3 correlation functions at imaginary frequency above critical temperature are shown on Fig.\,\ref{fig:imSigmaScan}, right panel. They show clearly that the poles corresponding to momentum relaxation $\Gamma$ (blue curve), the leading instability mode $\Psi_1$ (green mode) and the damped Goldstone mode $\Psi_2$ (yellow curve) appear separately in those 2-point function, meaning that their dynamics decouple. On the other hand below $T_c$ (left panel of Fig.\,\ref{fig:imSigmaScan}) we see that the $\Omega$ damped Goldstone mode leaves imprint in the  $\la J_x J_x \ra$ correlator, while $\Gamma$ is visible in $\la \Psi_{2} {J_y}_{2} \ra$, which signals that the two channels get coupled.

In order to carefully study the nature of the quasinormal modes, which get singular at the peaks of Fig.\ref{fig:imSigmaScan} we plot the profiles of the solutions to the linearized equations at frequencies corresponding to the peaks. These profiles are shown on Fig.\,\ref{fig:mode_profiles} and they agree very well with the interpretation of these modes as quarter period shifted structures, given in Sec.\,\ref{sec:model_TSB}.

In order to solve the numerical problems we use the 4-order nearest neighbor approximation of partial derivatives on the homogeneous grid of the size $20_x \times 40_z$. This resolution is adequate for our task, since we check that by increasing the resolution we don't observe any visible change in the positions of QNMs.
We use Wolfram Mathematica \cite{Mathematica} and the Python package developed in \cite{Balm:2019dxk}\footnote{A.K. thanks Floris Balm for help in setting up the software.} The numerical implementation if based on \cite{Krikun:2018ufr}.

\section{\label{app:helix}Bianchi VII helical model}

We consider the holographic model for translation symmetry breaking with the action \eqref{eq:action_helix}. The helical symmetry of the geometry allows us to reduce our calculations to solving systems of coupled ODEs, which greatly simplifies the calculations. 
For the background configuration, we take the ansatz for the metric and matter fields to be
\begin{align}
\label{ansatz helix 1}
 	ds^2 &= u^{-2} [ - T f  dt^2 + U/f du^2 + W_1 (\omega^{(p_c)}_1)^2 + W_2 (\omega^{(p_c)}_2 + Q dt )^2 + W_3 (\omega^{(p_c)}_3)^2] \\
 	A &= A_t dt + A_2 \omega^{(p_c)}_2 \\
 \label{ansatz helix 3}
 	B &= B_t dt + B_2 \omega^{(p_c)}_2
\end{align} 
\noindent where $\omega^{(p_c)}_i$ are the helical forms in \eqref{equ:helical_forms} and 
\begin{equation}
	f =  (1-u^2) (1 + u^2 -\mu ^2 u^4/3 )
\end{equation}
All unknowns are functions of the radial coordinate $u$ only.
We fix the gauge redundancy in our ansatz by imposing the DeTurk gauge as explained in \cite{Andrade:2017cnc}. 
The unbroken phase corresponds to Schwarzschild AdS in which 
\begin{align}
	T &= U = W_i = 1, & Q &= 0,  & A_t &= \mu (1- u^2), & A_2 &= B_t = B_2 = 0.
\end{align} 
Translations are broken spontaneously by a marginal mode of finite momentum, whose spatial dependence is fully encapsulated by the helical forms. At the linear level, the fields that are turned on in this marginal mode are $\delta g_{t2} \sim Q$ and $A_2$.  The breaking is spontaneous since the leading terms (corresponding to sources) in these fields are strictly zero. 
Choosing the CS coupling as $\bm{\gamma} = 3$, the marginal mode of highest temperature occurs at 
\begin{equation}
\label{equ:helix_preferred}
	p_c/\mu = 2.18, \qquad (T/\mu)_c = 0.223
\end{equation}
\noindent On top of the spontaneously broken phase, we introduce explicit breaking by choosing the boundary condition \eqref{eq:bc B2}, which in turn requires all the fields in the ansatz \eqref{ansatz helix 1}-\eqref{ansatz helix 3} to be non-vanishing. 

Another interesting perspective is to begin with backgrounds with explicit breaking only, set by the helical form \eqref{eq:bc B2} with pitch $k$, and study the marginal modes that trigger the spontaneous breaking around them. This study has been carried out previously in \cite{Andrade:2015iyf}. We now elucidate the existence of a secondary instability, not discussed in \cite{Andrade:2015iyf}, and which puts our model in the same footing as the general framework described in Section \ref{sec:model_TSB}. 

Let us describe this analysis here. The explicitly broken backgrounds have $Q = A_2 = B_t = 0$, with boundary condition \eqref{eq:bc B2}. We consider time-independent perturbations, which in addition to the spatial dependence present in the helices carry extra momentum~$q$. In this case, we obtain two sets of modes
\begin{equation}
 	(\delta g_{t2}, \delta A_2, \delta B_t)\propto e^{i q x}	, \qquad (\delta g_{t3}, \delta A_3) \propto e^{i q x},
\end{equation} 
\noindent which couple when $q \neq 0$. We further note that the mode $\delta B_t$
couples through a non-zero value of the background field $B_2$. Thus, and taking into account $\omega^{(k)}_{3}( x + \pi /(2 k) ) = \omega^{(k)}_{2}(x)$, we conclude that these two sets of modes are simply shifts from each other when $\lambda = 0$, and correspond to \eqref{equ:omega_modes}.
Stability requires the marginal modes to have real $q$. Therefore, we solve the eigenvalue problem given by the linear set of equations, and restrict to the set of real $q$ solutions (which may be empty). 

\begin{figure}
\center
\includegraphics[width=0.9 \linewidth]{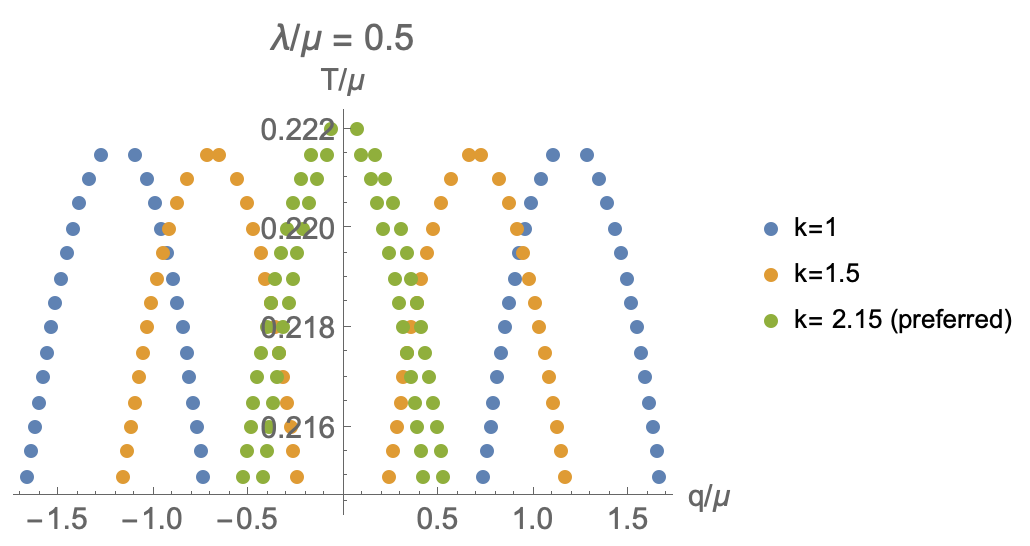}
\caption{\label{fig:helix_MM} 
\small
Marginal modes on the $(q,T)$ plane for different values of the helical pitch. Close to the preferred pitch 
$k \sim p_c = 2.18 \mu$ the primary and secondary instabilities lie on top of each other, which accounts for the structure of the QNMs in the broken phase, shown on Fig.\,\ref{fig:helix_spectrum}. All backgrounds have $\lambda/\mu = 0.5$.}
\end{figure}

We have studied the marginal modes for backgrounds with $\lambda/\mu = 0.5$ for varying temperature and selected values of background pitch $k$. 
As customary, we show the marginal modes in the $(q, T)$ plane, which results in the bell-shaped curves on
Fig.\,\ref{fig:helix_MM}. For $k/\mu = 2.15$, i.e. close to the preferred momentum \eqref{equ:helix_preferred}, we observe that both bell curves lie exactly on top of each other. The peaks correspond to the leading and subleading instabilities described in Section\,\ref{sec:model_TSB}. 
With critical
temperatures $T_1 = 0.2221$ and $T_2 = 0.2219$ they agree precisely with the results we obtain from the QNM analysis in Section\,\ref{sec:helical}. Indeed, on Fig.\,\ref{fig:helix_spectrum} we see that the leading instability, green mode, hits the axis at $T_1$, while the yellow mode, if extrapolated to the temperatures below $T_c$, would hit it at $T_2$.
We obtain qualitatively similar results for smaller values of $\lambda$. 
We stress that the leading instability is characterized by having the modes $(\delta g_{t2}, \delta A_2, \delta B_t)$ turned on, while for the subleading one we have $(\delta g_{t3}, \delta A_3)$ being non-zero. Note that this distinction can only be made at $q = 0$ where both sectors decouple. 

\vspace{0.3cm}
 
In order to study transport, we consider linear perturbations on top of the background configurations described above. For our purposes, it suffices to consider fields with harmonic time dependence of frequency $\omega$ (i.e. all fields are proportional to $\propto e^{- i \omega t}$), and the same helical structure as the background, i.e. the perturbations carry no extra momentum and $q=0$. 
We consider backgrounds connected to the primary instability at fixed pitch $p_c = 2.16 \mu$. On top of these, we turn on explicit breaking with $\lambda/\mu =  \{ 0.1, 0.2, 0.3, 0.4, 0.5 \}$, for which we find the critical temperatures $T/\mu = \{0.22325, 0.223, 0.22275, 0.2225,0.2221\}$, respectively. Using the formalism of \cite{Donos:2015gia, Donos:2015bxe} we have extracted the DC value for the conductivity, which we show in Fig.\,\ref{fig:helix_DC_at_Tc}.

We follow \cite{Andrade:2017cnc} to perform the computations for the conductivity and QNMs. In particular, we consider all possible perturbations and add gauge fixing terms which implement the DeDonder gauge for the metric and the Lorentz gauge for the vectors fields. We impose ingoing boundary conditions for the fluctuations at the horizon, and turn on the source corresponding to the leading term in $\delta A_1$ to extract the conductivity.  The QNMs, on the other hand, have zero sources in all linearized fields. We have carried out the numerical calculations for the background and fluctuations using a finite difference method of fourth order on a homogeneous grid with 160 points. After discretization, the full spectrum of QNMs can be obtained by solving the eigenvalue problem for the resulting system of linear algebraic equations. We have solved the resulting linear algebra in Mathematica \cite{Mathematica}. 

Above $T_c$, we find 4 decoupled sets of perturbations, %
\begin{align}
{\rm set} \, 1a &= \delta (A_1, g_{t1}, g_{u1}, g_{23}, B_3) \\
\notag
{\rm set} \, 1b &= \delta (A_3, g_{t3}, g_{u3}, g_{12}, B_1) \\ 
\notag
{\rm set} \, 2a &= \delta (A_2, g_{t2}, g_{u2}, g_{13}, B_t, B_u ) \\
\notag
{\rm set} \, 2b &= \delta (A_t, A_u, g_{uu}, g_{ut}, g_{11}, g_{22}, g_{33}, B_2)
\end{align}
\noindent Below $T_c$, set\,$1a$ with set\,$1b$, and set\,$2a$ with set\,$2b$, couple to each other, but modes denoted by $1$, $2$ remain decoupled.  
This structure provides interesting insight into the different physics that are controlled by the various modes in the spectrum as we now discuss. 

Recall that the optical conductivity is controlled by the mode $\delta A_1$. Above $T_c$, this is contained in the modes of set\,$1a$. The conductivity shows the usual Drude peak, which is linked to the fact that the only QNM in this sector that remains close to the origin is purely imaginary. We identify this mode as $\omega = - i \Gamma$, the value of which remains essentially constant across $T_c$, see Fig,\,\ref{fig:helix_spectrum}.
As stated above, the mode that governs the secondary instability is in set\,$1b$. We also find only one light QNM in this sector, which is purely imaginary. Above $T_c$ set\,$1b$ fully decouples from set\,$1a$, so this mode does not enter in the conductivity. However, as we decrease $T$ below $T_c$, these two sectors couple to each other, so that the conductivity receives contributions from two purely imaginary modes. This allows us to identify the light excitation in set\,$1b$ as $\omega = - i \Omega$. Further lowering the temperature, we find that the coupled $\tilde{\Gamma}$ and $\tilde{\Omega}$ modes collide and move off to the complex plane, giving rise to the pinned Goldstone mode. 

Turning our attention to set\,$2a$, we see that this is where the primary instability of the explicitly broken phase sits. We find that 
this sector contains only one light QNM which is purely imaginary, and corresponds to the ``bouncing mode'' (green on Fig.\,\ref{fig:helix_spectrum}), which stabilizes once the order parameter develops below $T_c$. Due to the decoupling of modes, this excitation does not influence the optical conductivity. Similarly, the modes in set\,$2b$ decouple from set\,$1$. At $q = 0$ these modes only contribute to the set of light modes with one $\omega = 0$ excitation, which is related to the standard sound mode of AdS branes (i.e. corresponds to fluctuations of the energy density and pressure).  

This intricate mode dynamics can be captured by the formula \eqref{AC model}, which remains valid in the whole range of parameters we have explored. In particular, we note that as long as there are two modes or fewer near the real axis in set\,$1$, the conductivity can be approximated by \eqref{AC model}. This holds true across the 3 regimes in temperature: 

\begin{itemize}
\setlength\itemsep{0em}

\item Above $T_c$: There is one QNM in set\,$1a$: $\omega = - i \Gamma$, and the conductivity displays a Drude peak of this width. 
The mode $\omega = - i \Omega$ in set\,$1b$ decouples. 

\item Below, near  $T_c$: there are two purely imaginary modes in set\,$1$: $\omega = - i \tilde{\Gamma}$, $\omega = - i \tilde{\Omega}$. There is an off-center bump in the conductivity, shown on Fig.\,\ref{fig:helix_AC}. 

\item Below, further from  $T_c$: the imaginary modes have collided and produced a pair of complex conjugate QNMs which we identify as the pinned Goldstone. The conductivity shows a sharp peak whose location is given by the real part of the QNMs, see results of \cite{Andrade:2017cnc}.

\end{itemize}

We have mainly focused in the range of temperatures near $T_c$, where $\omega_0$ is small. This makes the contributions 
from the dissipative parameter $\gamma$ in \eqref{AC model} negligible. In this regime our numerical data can be well described by \eqref{AC model} with $\gamma = 0$. 

\section{\label{app:TDGL}TDGL equation with no residual $\mathbb{Z}_2$ symmetry}

Let us consider here the case when the external source breaks the global symmetry completely, as it is the case in some homogeneous Q-lattices and linear axion models. In this case there is no residual $\mathbb{Z}_2$-symmetry remaining in the normal phase and therefore no spontaneous symmetry breaking happens at $T_c$. Strictly speaking in this situation one can not apply the notions of second order phase transition, order parameter, etc. However if the external source is small, then the phenomenology is very reminiscent to the phase transition and therefore such cases may be called ``imperfect phase transition'' \cite{golubitsky1979imperfect}. In case of conventional periodic patterns of TSB this corresponds to a $n=1/1$ commensurate state, where, unlike the $n=2/1$ case discussed in the main text, no residual discrete shifts of the spontaneous structure are allowed after the explicit potential is turned on \cite{PhysRevLett.56.724}, see our discussion preceding \eqref{equ:amplitude equation_deformed}. The corresponding symmetry breaking term in the free energy in case of $n=1/1$  reads (c.f.\eqref{equ:GL_deformed})

\begin{equation}
%\label{equ:GL_deformed}
\mathcal{F}_f = \alpha |\Phi|^2 + \frac{\beta}{2} |\Phi|^4 - f^* \Phi - f \Phi^*.
\end{equation}
(Note that the $\Phi \rar - \Phi$ symmetry is lost)
Therefore the TDGL equation takes the form
\begin{equation}
%\label{equ:def_time_GL}
\p_t \Phi = -\frac{\delta \mathcal{F}}{\delta \Phi*} = - \alpha \Phi - \beta |\Phi|^2 \Phi + f.
\end{equation}
Without loss of generality we can choose $f$ to be real and positive. The possible ground states of the system above and below $T_c$ are now given by the real solutions to the cubic equation $\p_t \Phi = 0$. According to the algebra of cubic equations one real root is always present. Note however, that $\Phi=0$ is not a solution anymore even above $T_c$, so formally the order parameter is always finite.

\begin{figure}
\center
\includegraphics[width=1 \linewidth]{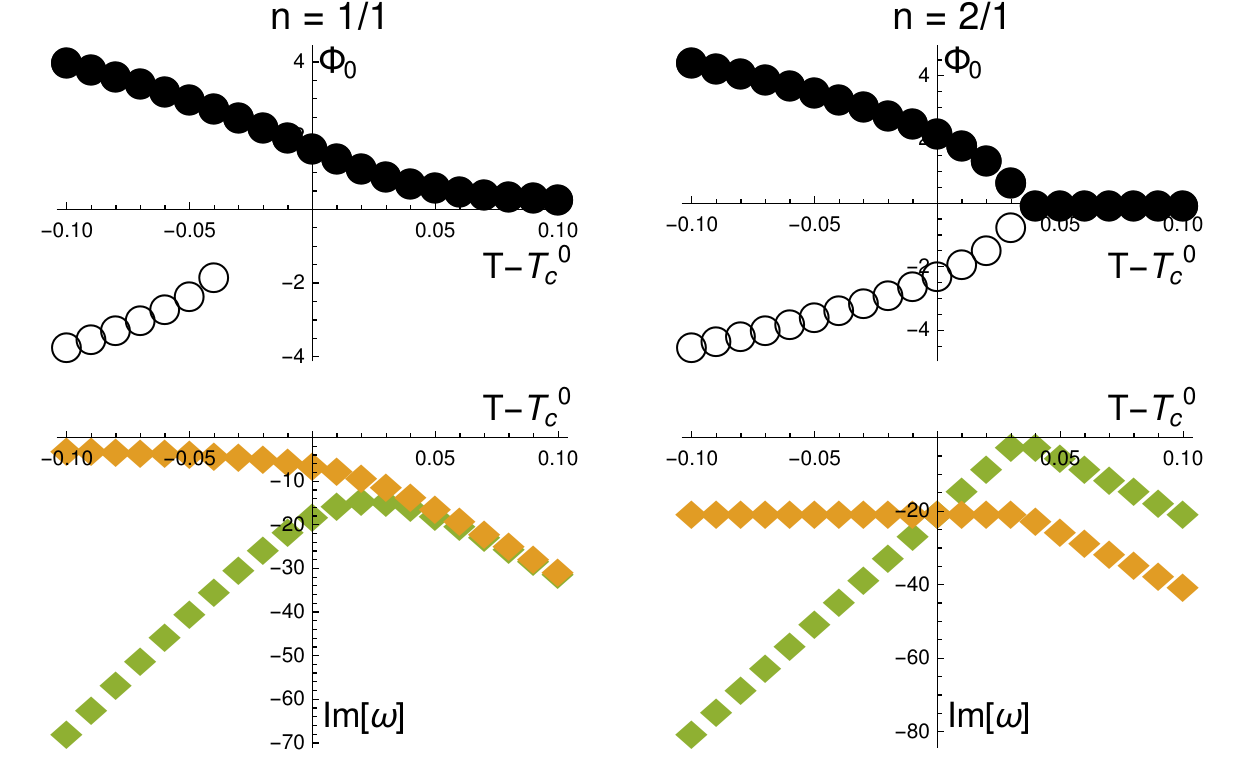}
\caption{\label{fig:TDGLE} 
\small
\textbf{Comparison of the cases with and without residual $\mathbb{Z}_2$ symmetry.} The left column corresponds to $1/1$-commensurate case in terms of \cite{PhysRevLett.56.724}, where no residual symmetry is present. The right column depicts $2/1$-commensurate case with spontaneous residual  $\mathbb{Z}_2$ symmetry breaking at critical temperature. \\
\textbf{The top row} shows the ground states of the system depending on temperature ($T_c^0$ is the critical temperature in absence of explicit source). The black dots show the globally thermodynamically preferred solutions, while the empty circles -- the quasistable local minima of the free energy. In $1/1$-case there is no well-defined phase transition since the value of $\Phi_0$ is always finite, while in $2/1$-case $\Phi_0$ plays the role of the order parameter. Note that the phase transition happens above $T_c^0$ due to the effect of explicit source, c.f. \eqref{equ:deformed_modes_above_Tc}. \\
\textbf{The bottom row} Shows the spectra of quasinormal modes, corresponding the fluctuations of the real (green) and imaginary (yellow) components of the scalar field. (Same as Fig.\,\ref{fig:U1QNMs}) Note that in $1/1$ case there is no instability mode hitting the real axis, like in $2/1$-case. Importantly for us, however, the behavior of the phase relaxation mode below $T_c^0$ is similar in both cases: it stays close to the real axis with finite damping proportional to explicit scale $f$.
}
\end{figure}

The two extra real solutions appear when the discriminant
\begin{equation}
\Delta = - \left(\frac{\alpha}{3 \beta}\right)^3 - \left(\frac{f}{2 \beta}\right)^2
\end{equation}
gets positive. Similarly to the $n=2/1$ case, the temperature $T_c^f$, when this happens, differ from $T_c$ due to finite $f$. Below $T_c^f$, however, the solution of the ``normal phase'' smoothly connects to the one of the branches of solutions in the ``broken phase'', so no bifurcation occurs, which would be associated with residual symmetry breaking in case of $n=2/1$. To illustrate this, we compare the two cases for particular numerical parameters: $a = 300 (T-T_c), T_c=0, f= 10, \beta = 2$ on the top row of Fig.\,\ref{fig:TDGLE}. The leading solution is shown with black dots, while the metastable one as empty circles.

Going further, we analyze the stability of the ground states in complete analogy with \eqref{equ:def_modes_below_Tc}, by studying the fluctuations of real and imaginary parts of the scalar. The corresponding equations for $n=1/1$ case read (note the absence of $f$-term)
\begin{align}
\p_t \delta \phi_1 &= - \left( \alpha + 3 \beta \Phi_0^2 \right) \delta \phi_1, \\
\notag
\p_t \delta \phi_2 &= - \left( \alpha + \beta \Phi_0^2 \right) \delta \phi_2.
\end{align}
The analytic expressions for $\Phi_0$ in case of the cubic equations are cumbersome, so we don't get the same beautiful results as in $n=2/1$ case. However, as one can see from our numerical examples on the bottom row of Fig.\,\ref{fig:TDGLE}, the qualitative behavior of the QNMs is roughly the same: one of them -- the amplitude mode -- departs to the lower complex half-plane below $T_c$, while the other one -- the phason -- keeps close to the real axis, with finite damping due to the explicit source $f$. This explains why for homogeneous models discussed in Sec.\,\ref{sec:homogeneous} we get qualitatively the same results for phase relaxation $\Omega$, as for the other cases with $n=2/1$ commensurability and spontaneous breaking of residual $\mathbb{Z}_2$.

\bibliographystyle{JHEP-2}
\bibliography{omega_Tc}

\end{document}